\documentclass[10pt,conference]{IEEEtran}

\usepackage{mathptmx} 
\usepackage{tcolorbox}
\usepackage{amsfonts}
\usepackage{subcaption}
\usepackage{fancyhdr}
\usepackage[normalem]{ulem}
\usepackage[hyphens]{url}
\usepackage[sort,nocompress]{cite}
\usepackage[final]{microtype}
\usepackage[keeplastbox]{flushend}
\usepackage{multirow}
\usepackage{graphicx}
\usepackage{ragged2e}
\usepackage{soul}
\usepackage[flushleft]{threeparttable}
\usepackage{breakurl}

\usepackage[bookmarks=true,breaklinks=true,letterpaper=true,colorlinks,linkcolor=black,citecolor=blue,urlcolor=black]{hyperref}
\usepackage{comment}
\usepackage{listings}
\def\BibTeX{{\rm B\kern-.05em{\sc i\kern-.025em b}\kern-.08em
    T\kern-.1667em\lower.7ex\hbox{E}\kern-.125emX}}
\usepackage{array}
\newcolumntype{P}[1]{>{\raggedright\arraybackslash}p{#1}}

\makeatletter
\AtBeginDocument{\let\hl\@firstofone}
\makeatother

\newcommand{\chl}[1] {\textcolor{black}{#1}}
\newcommand{\fhl}[1] {\textcolor{black}{#1}}
\newcommand{\lhl}[1] {\textcolor{black}{#1}}

\usepackage{array}
\usepackage{hyperref}
\usepackage{makecell}
\newcolumntype{C}[1]{>{\centering\arraybackslash}m{#1}}
\usepackage{tikz}

\newcommand{\acronym}{AutoCAT}
\title{\acronym: Reinforcement Learning for Automated Exploration of Cache-Timing Attacks}

\author{
\IEEEauthorblockN{Mulong Luo\IEEEauthorrefmark{1}\textsuperscript{1}, 
Wenjie Xiong\IEEEauthorrefmark{2}\IEEEauthorrefmark{3}\textsuperscript{1}, 
Geunbae Lee\IEEEauthorrefmark{2},
Yueying Li\IEEEauthorrefmark{1}, 
Xiaomeng Yang\IEEEauthorrefmark{3},\\
Amy Zhang\IEEEauthorrefmark{3},
Yuandong Tian\IEEEauthorrefmark{3},
Hsien-Hsin S. Lee\IEEEauthorrefmark{4}$^2$,
and G. Edward Suh\IEEEauthorrefmark{1}\IEEEauthorrefmark{3}}\\
\IEEEauthorblockA{\IEEEauthorrefmark{1}Cornell University~~\IEEEauthorrefmark{2}Virginia Tech~~
\IEEEauthorrefmark{4}Intel Corporation~~
\IEEEauthorrefmark{3}Meta AI}
\IEEEauthorblockA{\{ml2558,yl3469,gs272\}@cornell.edu, \{wenjiex,geunbae\}@vt.edu, linear@acm.org,}
\IEEEauthorblockA{\{wenjiex,yangxm,amyzhang,yuandong,edsuh\}@meta.com }
}

\begin{document}
\maketitle
\begingroup\renewcommand\thefootnote{1}
\footnotetext{Equal contributions.}
\begingroup\renewcommand\thefootnote{2}
\footnotetext{Work done while at Meta AI.}
\endgroup

\pagestyle{plain}
\pagestyle{empty}

\begin{abstract}
The aggressive performance optimizations in modern microprocessors can result in security vulnerabilities. 
For example, timing-based attacks in processor caches can steal secret keys or \lhl{break randomization}. 
So far, finding cache-timing vulnerabilities is mostly performed by human
experts, which is inefficient and laborious. There is a need for automatic tools that can explore vulnerabilities given that unreported vulnerabilities leave the systems at risk.

In this paper, we propose \textit{\acronym}, an automated exploration framework that finds cache timing-channel attack sequences using reinforcement learning (RL). 
Specifically, \acronym{} formulates the cache timing-channel attack as a guessing game between an attack program and a victim program holding a secret. This guessing game can thus be solved via modern deep RL techniques.
\acronym{} can explore attacks in various cache configurations without knowing design details and under different attack and victim \lhl{program} configurations. \acronym{}  can also find attacks to bypass certain detection and defense mechanisms. In particular, {\acronym} discovered \emph{StealthyStreamline}, a new attack that is able to bypass \lhl{performance counter-based detection} and has \hl{up to a $71\%$} higher information leakage rate than the state-of-the-art LRU-based attacks \hl{on real processors}. {\acronym} is the first of its kind \lhl{in} using RL for crafting microarchitectural timing-channel attack sequences and can accelerate cache timing-channel exploration for secure microprocessor designs.

\end{abstract}

\section{Introduction}
\label{sec:intro}

%
%
As we use computers to handle increasingly sensitive data and tasks,
security has become one of the major design considerations for modern computer systems.
For example, from the hardware perspective, microarchitecture-level timing channels 
have emerged as a major security concern as they allow leaking of information covertly with a high bit-rate and 
bypassing the traditional software isolation mechanisms.
The timing-channel attacks are also shown to be an even more serious problem when combined with
speculative execution capabilities \cite{lipp2018meltdown,kocher2019spectre}.

%
%
Unfortunately, developing a system that is sufficiently secure and \lhl{performant} at the same time 
is quite challenging in large part because it is difficult to evaluate the security of a system design.
By definition, a security vulnerability comes from an unknown bug or unintended use of a system feature, 
which is difficult to know or quantify at design time.
While formal methods and cryptography can provide mathematical guarantees, it is difficult to scale
the formal proofs to complex systems and security properties in practice.
As a result, today's security evaluations and analyses largely rely on human reviews
and empirical studies based on known attacks or randomized tests. 
However, the security evaluation based on known attack sequences manually discovered by humans makes it difficult
to assess the security of a new microarchitecture design or a defense mechanism. 
Vulnerabilities in new microarchitectures are often left unnoticed for a long time, and defense mechanisms are
often found vulnerable to new attack sequences even when they are similar to the known ones.
For example, while caches existed in microprocessors for a long time, Bernstein's cache-timing attack 
was reported back in 2005 \cite{bernsteincache} followed by multiple variations such as 
evict+time (2006) \cite{osvik2006cache}, flush+reload (2014) \cite{yarom2014flush+}, flush+flush (2016) \cite{gruss2016flush+}, etc. New attacks in caches are still being reported recently, e.g., attacks in cache replacement states (2020)~\cite{xiong2020leaking,briongos2020reload+}, streamline (2021)\cite{saileshwar2021streamline}, and attacks using cache dirty states (2022)~\cite{cui2022abusing}. 

%
%
This paper proposes to leverage reinforcement learning~(RL) to automatically explore attack sequences for microarchitectural timing-channel vulnerabilities, 
and demonstrate the feasibility of this approach using cache timing channels as a concrete example. RL has been shown to achieve super-human performance in multiple competitive games (e.g., Go and Chess~\cite{silver2017mastering}, DoTA 2~\cite{berner2019dota}) without starting from \lhl{strategies based on human experience}. In this paper, we show that microarchitecture-level timing-channel attack can also be formulated as a \emph{guessing game} for an attack program and is a good fit for RL. In this case, an RL agent can learn by self-playing the game many times in a well-defined environment, which is provided by real hardware or efficient simulation infrastructures commonly used for architecture studies.
The experiments show that our RL framework, named \emph{\acronym}, can automatically adapt to a variety of cache designs and countermeasures, and find cache-timing attacks, including known attack sequences and ones that are more efficient than known attack sequences. 

%
%
While the use of machine learning (ML) for system security has been explored in the past, 
the previous work largely focused on performing or detecting {\em known attack sequences} on known system designs.
For example, ML models with supervised learning are used in side-channel attacks to recover secrets \cite{michalevsky2015powerspy,yan2019attack,la2021wireless,wei2018know,yuan2022automated,luo2020stealthy}.
Similarly, ML models can be trained with known attack traces 
for intrusion detection \cite{lane2000machine}.
This paper asks a different question: can an RL agent automatically
1)~learn system designs without explicit specifications and 2)~generate attack sequences that are not specified by humans? 
\hl{To be widely applicable, the RL agent should also be able to adapt to diverse new system designs without substantial changes to the RL environment.}
Our experimental results suggest that such autonomous explorations are indeed possible.

%
%
We believe that the RL-based approach has the potential to enable a more
systematic and rigorous evaluation of system security. 
We envision the RL framework to be used for both 1)~studying potential 
security vulnerabilities of a system design and 2)~evaluating the robustness of a defense mechanism.  
For example, {\acronym} can be used to automatically generate cache-timing attack sequences for
a diverse set of cache configurations, replacement policies, or real processors.
{\acronym} also tries to find attack sequences with higher success rates and bandwidth, providing
a way to quantitatively compare the effectiveness of attacks across different cache designs.
\acronym's environment can be augmented with an explicit protection scheme such as an attack detector,
and the RL agent can be asked to find an attack sequence that bypasses the defense.
While {\acronym} cannot prove the security of a defense mechanism, testing a countermeasure with {\acronym} will provide a better measure of its robustness compared to only testing its effectiveness using 
the known attack sequences that it is designed for.

%
%

%
%

%
%
The following summarizes the main technical contributions and experimental findings of this paper:
\begin{itemize}
  \setlength{\itemsep}{0pt}
  \setlength{\parskip}{0pt}
  \setlength{\parsep}{0pt}
    \item We present \acronym, the first framework to use RL to automatically explore cache-timing attacks. The framework can interface with  a cache simulator or a real processor.
    \item We demonstrate that {\acronym} can find cache attack sequences for multiple cache configurations, replacement policies, and \hl{prefetchers}. \acronym{} can also find attack sequences on multiple real processors with unknown replacement policies quickly, while manually applying known attack strategies to a new processor design requires significant reverse-engineering. 
    
    \item We demonstrate that {\acronym} can bypass several cache-timing defense and detection schemes, such as the partition-locked (PL) cache~\cite{wang2007new}, detection based on the victim \lhl{program} misses~\cite{zhang2016cloudradar,chiappetta2016real,alam2017performance,kulah2019spydetector,mirbagher2020perspectron}, detection based on autocorrelation~\cite{chen2014cc,yan2016replayconfusion}, and \hl{ML-based detectors}~\cite{harris2019cyclone}.
    
    \item We present a novel cache-timing attack, named {\em StealthStreamline}, discovered by \acronym, which avoids detection based on the miss counts and has \hl{an up to $71\%$} higher bit rate than existing LRU-based attacks \hl{on real machines}.
\end{itemize}

%
%
The rest of the paper is organized as follows. 
Section~\ref{sec:background} discusses background and motivation.
Section~\ref{sec:overview} provides the high-level overview of our methodology.
Section~\ref{sec:design} describes the design and implementation of \acronym.
Section~\ref{sec:case_studies} presents the experimental results, including several case studies.
Section~\ref{sec:discussions} discusses other aspects of \acronym.
Section~\ref{sec:related_work} discusses related work, and
Section~\ref{sec:conclusion} concludes the paper. The code is made publicly available at \url{https://github.com/facebookresearch/AutoCAT}.


\section{Background and Motivation}

\label{sec:background}


\subsection{Cache-Timing Attacks}
\label{sec:background_cache}


The cache timing channel is a widely-studied vulnerability in modern microprocessors \lhl{because of} its practicality and high bit rate. Depending on the threat model, it can be used as a side channel (where an attack \lhl{program} and a victim \lhl{program}  are non-cooperative) or a covert channel (where a sender and a receiver cooperate) on a shared cache to steal or send information. Without loss of generality, we assume a side channel scenario.

The cache-timing channels usually involve two parties, the attack \lhl{program} and the victim \lhl{program}. 
The victim \lhl{program}  has a secret, and the memory operation of the victim \lhl{program} depends on the secret.
The attack program's goal is to guess the secret without directly accessing the secret. 
The attack program needs to achieve this goal by making memory accesses and measuring the timing of the memory accesses or the victim \lhl{program}'s external activities, e.g.,  using the timing/cycle measurement facilities such as \texttt{RDTSCP} in x86. 
The timing of memory access indicates whether a cache line is in the cache or not. 
For example, in a prime+probe attack, the victim \lhl{program}'s memory access will evict the attack program's cache line from the cache, causing latency changes in the attack program's future memory accesses. Thus, the attack program can infer the victim \lhl{program}'s memory operation from timing observations.

The recent cache timing-channel models \cite{deng2020benchmark,weber2021osiris} divide the attacks into  
three essential components, including the attack program's actions, the victim \lhl{program}'s actions, and the attack program's observations.
The attack program's actions include normal memory accesses, cache line flushing, etc. The victim \lhl{program}'s action is usually a secret-dependent memory operation such that it changes the state of the cache, e.g., accessing a secret address ($addr_{secret}$).
For the attack program's observations, the attack program accesses certain cache lines and obtains the timing measurement to gain information about the secret ($addr_{secret}$). 
With correct combinations, the attack program can learn the secret, as demonstrated in known attacks such as prime+probe \cite{liu2015last}, flush+reload~\cite{yarom2014flush+}, evict+time \cite{osvik2006cache}, cache collision \cite{bonneau2006cache} attacks, and other recent cache timing-channel attacks. 
Table~\ref{tab:three_step} summarizes the actions of the common cache timing-channel attack categories~\cite{he2017secure,deng2020benchmark}.
Other cache-timing attacks also share the same set of actions and observations.

In this paper, we refer to a sequence of actions such as memory accesses, cache flushing, allowing a  victim program's execution, and others in an attack on a specific system as an attack sequence. 
An attack category or strategy is a broader class of attack sequences that can be adapted to multiple systems, similar to the ones in Table~\ref{tab:three_step}.

\subsection{Challenges of Cache-Timing Attack}

For security analysis and testing, given a hardware design, we need to analyze the vulnerabilities and generate attack sequences to exploit these vulnerabilities. Many attack sequences may belong to a previously known attack category, while there may also be new attack sequences. Even for known attacks, there is a need to adapt those to the given design to test the effectiveness and bit rate of the attack~\cite{wang2019papp}. In general, a cache timing attack involves (1) reverse-engineering the behavior of target micro-architecture design, (2) designing attack sequences to expose information, and (3) improving the signal quality. 

{\acronym} aims to address the challenges in the first two steps by 
focusing on reverse-engineering and attack sequence exploration when new or blackbox hardware is given.
For other practical challenges in developing end-to-end attacks, such as building an accurate timer, reducing background noise, and synchronizing the  victim \lhl{program}, 
we leverage the existing techniques in the  literature~\cite{vila2020cachequery,xiong2020leaking,yan2019attack,saileshwar2021streamline}.
The following discusses some of the main challenges in the attack steps. 

 A cache contains many design options, such as a replacement policy\cite{perez2011functional}, cache directory \cite{yan2019attack}, \hl{cache prefetcher}~\cite{wang2019papp,chen2021leaking}, etc. The microarchitectural implementation details are often kept as proprietary information by chip designers. Cache operations can also be pseudo-random such as random replacement policies, meaning the cache behavior is not entirely predictable. Reverse engineering effort can be onerous and only reveals limited information~\cite{abel2014reverse,vila2020cachequery}. For example, it could take up to 100 hours to reverse-engineer a replacement policy~\cite{vila2020cachequery}.

\begin{table}[t]
\centering
\caption{Actions/observations in known cache timing attacks.}
\label{tab:three_step}
\begin{tabular}{|@{ }c@{ }|@{ } c@{ }|@{ } c@{ }|@{ }c@{ }|} 
 \hline
 {\bf Attack Category} & {\bf Attacker  actions} & {\bf Victim  actions} & {\bf Observations} \\
\hline
prime+probe~\cite{liu2015last}&  access addrs& access an addr& attacker's  latency\\
\hline
flush+reload~\cite{yarom2014flush+} & flush  addrs& access an addr& attacker's  latency \\
\hline 
evict+reload~\cite{osvik2006cache} & access addrs & access an addr& attacker's  latency \\
\hline
evict+time~\cite{bonneau2006cache} & access addrs & access addresses &  victim's latency \\
\hline
\end{tabular}
\end{table}

Some of the microarchitecture states are not directly measurable by timing. The attack has to encode and decode the secret in such states, potentially resulting in complex and long attack sequences. For example, known attacks that use the cache replacement states and cache coherent states all require complex sequences of actions~\cite{xiong2020leaking,briongos2020reload+,cui2022abusing,yao2018coherence}, making it challenging to manually reason about  an exploitable vulnerability and design an attack sequence.

To explore attacks in a given design, the above process needs to be repeated, which is laborious, and thus, tools helping attack explorations have been proposed. 
One method is to manually model the existing attacks in the cache~\cite{deng2020benchmark}, and then automatically generate attacks based on the model. 
However, so far,  due to the complexity, such manual modeling methods~\cite{he2017secure,deng2020benchmark} are still limited to only the cache tag states, not including other cache states that could be vulnerable.
Rigorous approaches can determine if certain security exploits exist in the design. 
However, formal methods \cite{trippel2018checkmate}, and information flow tracking \cite{zhang2015hardware} require whitebox modeling of the design, which in many cases is hard or impossible for a commercial microprocessor. 
In addition, lifting a new RTL-level design into a formal \lhl{specification} without manual effort is still challenging \cite{hsiao2021synthesizing}. 
To deal with blackbox designs, fuzzing \cite{weber2021osiris,ghaniyoun2021introspectre} has been used to automatically generate attack sequences. However, to bound the size of the search space, fuzzing usually requires pre-defined attack sequences or predefined gadgets, which limit the attack strategies that can be explored. 
Ideally, we want fewer restrictions on the form of attack sequences and a tool that can explore (both known and unknown) attack sequences in blackbox designs.


\subsection{Reinforcement Learning (RL)}
\label{sec:background_RL}

RL aims to find a policy that generates an action sequence that maximizes long-term rewards. Figure~\ref{fig:RL} shows the high-level components and concepts in RL. First, there is an {\em RL agent}, which is controlled by a {\em policy}. There is an {\em environment}, which the RL agent interacts with. In each step, the RL agent takes an {\em action}, which feeds into the environment. The environment changes the state with the action, then exposes observations to the RL agent, and assigns {\em reward} values to the RL agent.  The environment can optionally have a state where the environment state is reset. We call the sequence between two adjacent resets an {\em episode}. The goal of RL agent training is to generate a sequence of actions within an episode so that the sum of the rewards within one episode is maximized.

\begin{figure}[t]
\centering
\includegraphics[width=0.6\columnwidth]{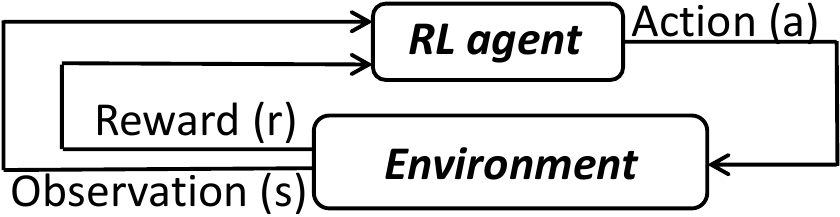}
\caption{RL formulation. RL agent does not need knowledge of the internals of the environment to learn a policy.}
\label{fig:RL}
\end{figure}

\begin{figure*}[t]
\centering
\includegraphics[width=1.9\columnwidth]{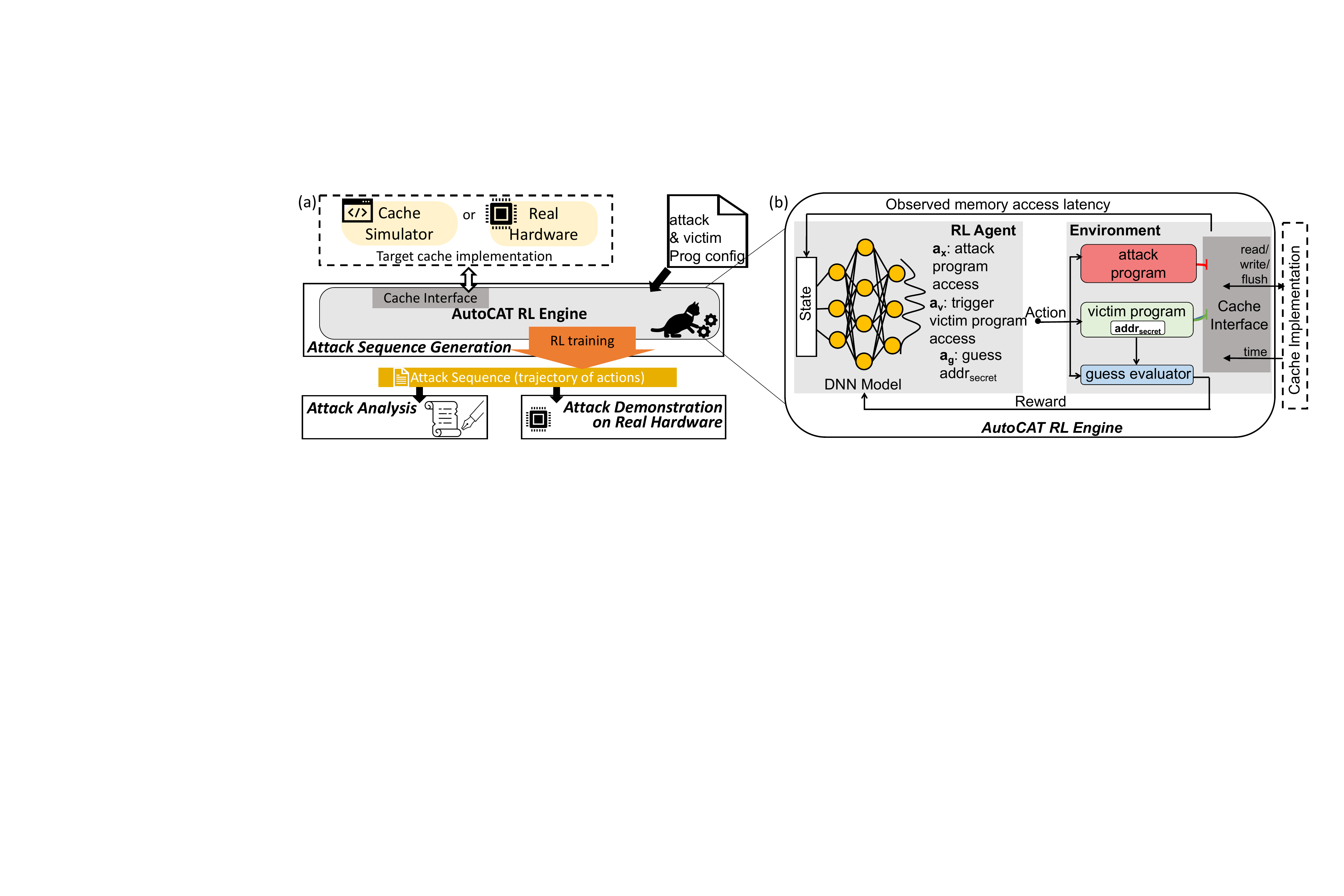}
\caption{(a) \acronym~framework. (b) \acronym~RL engine.
}
\vspace{-6pt}
\label{fig:RL_formulation}
\end{figure*}

Recently, RL has been shown to learn human-level (or even super-human-level) policies in many game environments~\cite{mnih_human-level_2015,silver_mastering_2016} where there are well-defined rules and win/loss conditions. Those games are relatively lightweight and can be simulated easily. 
RL even discovers novel tactics, including novel openings in the game of Go~\cite{silver2017mastering}, which humans have played for thousands of years. 
Similarly, cache-timing attacks can be formulated as a {\em guessing game}, where an attack program aims to guess a victim \lhl{program}'s secret correctly while paying a small penalty (negative reward) for each cache access before making a guess. 

The RL-based cache-timing attack discovery has several advantages over the alternatives.
First, the RL formulation imposes few restrictions on the length of an attack, allowing flexible exploration of a broader set of attack sequences compared to the existing tool~\cite{weber2021osiris}. Second, RL is agnostic to the implementation of the environment as long as the interface including the action, reward, and observation is provided. Thus, RL can work with both existing simulators as well as commercial processors without finding a formal white-box model of their caches~\cite{trippel2018checkmate}.
Third, the RL is parameterized with neural networks, which can generalize to unexplored states and make smart decisions \cite{lazaric2012transfer,narvekar2020curriculum}, compared to random test-case generations that need to explore each state independently. 
Thus, RL can be more efficient in searching for vulnerabilities.

\section{{\acronym} Overview}
\label{sec:overview}

\subsection{Overview of AutoCAT Framework}

Figure~\ref{fig:RL_formulation}(a) shows the \acronym's overall framework. \acronym's RL engine \lhl{takes a target cache implementation, the configuration of the attack program and the victim program, and the RL configuration to generate attack sequences.}
The cache implementation can be either a simulator (with a certain cache configuration) or a real hardware processor. 
The attack sequence is the sequence of memory operations, which can then be used in attack analysis by human experts to classify and identify potential new attacks. The attack sequence from the RL engine can also be used in an attack demonstration on real hardware. 
In this study, we manually analyzed the attack sequences to categorize them and applied a new attack sequence from {\acronym} to multiple Intel processors for real-world demonstration.

\subsection{High-level RL Engine Formulation}\label{sec:rl_engine}

Figure~\ref{fig:RL_formulation}(b) shows the formulation of cache guessing game as an RL problem  in \lhl{the} \acronym~RL engine. In this setting, we let the RL agent explore possible attack strategies by controlling the attack \lhl{program}. 
For simplicity, {\acronym} currently allows the RL agent to also control when the victim \lhl{program} runs.
The RL agent is \lhl{controlled by} a policy parameterized by a deep neural network (DNN) model. The environment interacts with a cache implementation, which handles the memory accesses of both the victim \lhl{program} and the attack \lhl{program}. The environment also contains $addr_{secret}$ which is the secret of the victim, and a guess evaluator to check if a guess is correct. 
Below we explain the RL formulation.


\noindent \textbf{Episode:} In each episode, the environment randomly generates the victim \lhl{program}’s secret address ($addr_{secret}$).
The agent can then explore different actions and get observations. When the agent decides to make a guess on $addr_{secret}$, the episode will terminate. Reward will be given based on the guessing result and the number of actions taken. 

\noindent \textbf{Rewards:} 
As the goal is to let the agent learn how to guess $addr_{secret}$ correctly, we set the environment to return a positive reward if the agent's guess is correct, and a negative reward if the agent makes a wrong guess. 
To encourage the agent to optimize the attack strategy (i.e., minimize the number of steps), we give a small penalty for each step the agent takes.

\noindent \textbf{Actions:}\begin{itemize}
  \setlength{\itemsep}{0pt}
  \setlength{\parskip}{0pt}
  \setlength{\parsep}{0pt}
  \vspace{-3pt}
\item $a_X$---access X where X is the memory address accessible by the attack \lhl{program}. As an attacker, the agent can access a cache line of address X, and observe a hit/miss.
\item $a_v$---trigger the victim \lhl{program}. The RL agent can trigger the victim \lhl{program}'s secret access; the victim \lhl{program}  accesses $addr_{secret}$, which potentially changes the cache state.
\item $a_{gY}$---guess that the value of  $addr_{secret}$ is $Y$ where Y is chosen from the addresses accessible by the victim \lhl{program}, and end the current episode \lhl{if sending one secret, or change the value of $addr_{secret}$ if sending multiple secrets in one episode, as shown in Autocorrelation and ML-based detection in Section~\ref{sec:bypass}}. 
\end{itemize}
\noindent \textbf{Observations:}
	When the agent takes $a_X$ (access X) above, the cache implementation conducts the memory operation, i.e., looks up the address X in the cache, returns the latency of the access, and updates the cache state. For $a_v$ (trigger victim \lhl{program}), the environment uses the $addr_{secret}$ of the current episode and lets the cache simulator access $addr_{secret}$.
	With this environment, the agent can explore attacks using a combination of actions.

\section{Design and Implementation}
\label{sec:design}

\begin{table*}[h!]
\centering
\vspace{+4pt}
\caption{{\acronym} configuration parameters.}
\label{tab:config}

\begin{tabular}{|l|l| p{6.75cm}| c|c|} 
 \hline
\bf Option Type&\bf Option Name & \bf Definition & \bf Type & \bf Range \\ 
 \hline
\multirow{3}{*}{Cache configs}   & \texttt{num\_blocks} &total number of blocks in the cache& integer & 2,4,...\\\cline{2-5}
   & \texttt{num\_ways} & number of ways of the cache&  integer & 2,4,6,8,12,16\\\cline{2-5}   in cache simulator & \texttt{rep\_\lhl{policy}} &replacement \lhl{policy} for all cache sets & str& "lru", "plru",... \\\hline

  & \texttt{attack\_addr\_s} & starting address of the attack program & integer& 0,1,2,3,... \\\cline{2-5}
   & \texttt{attack\_addr\_e} & end address of the attack program& integer & 0,1,2,3 ,...\\ \cline{2-5}
Attack and  & \texttt{victim\_addr\_s} &starting address of the victim program & integer& 0,1,2,3,... \\ \cline{2-5}
 victim program & \texttt{victim\_addr\_e} & end address of the victim program & integer & 0,1,2,3,... \\ \cline{2-5}
 configurations  &  \texttt{flush\_enable} &whether to enable flush instruction for the attack program &boolean& true, false \\ \cline{2-5}
   
    &  \texttt{victim\_no\_access\_enable} & whether the victim needs to make a memory access when triggered by the attack program & boolean & true,false \\\cline{2-5}
    &\texttt{detection\_enable} & whether the episode terminates when the attack detector signals a potential attack & boolean&true,false \\\hline
 \multirow{7}{*}{RL config}  & \texttt{window\_size} &size of the history window in the observation space & integer& 1,2,3 ,...\\\cline{2-5}
       
  &\texttt{correct\_guess\_reward} &reward when the attack program  makes a guess and the guess is correct&float & $(0,\infty)$ \\ \cline{2-5}
    &\texttt{wrong\_guess\_reward} & reward when the attack program makes a guess and the guess is wrong&float & $(-\infty, 0]$ \\ \cline{2-5}
    &\texttt{step\_reward} & reward when the attack program make a memory access &float&$(-\infty, 0]$ \\\cline{2-5}

    &\texttt{length\_violation\_reward} & reward value when the length of episode exceeds limit & float & $(-\infty,0]$  \\\cline{2-5}
    &\texttt{detection\_reward} &reward when the attack detector signals a potential attack & float & $(-\infty,0]$\\\hline

\end{tabular}
\vspace{-12pt}
\end{table*}

\subsection{Cache Implementation and Configuration}
\label{sec:cache_sim}

As depicted in Figure~\ref{fig:RL_formulation}, there are two choices for cache implementation: a cache simulator written in software or a real processor.
The cache simulator allows quickly prototyping and implementing existing and new cache designs for vulnerability exploration. 
Exploring attacks on real hardware enables applying {\acronym} to real-world system designs even when their design details are not known. 

\noindent{\bf Cache simulator}:
we embed an \hl{open-source} cache simulator in Python~\cite{cache_simulator}  as the cache model in the {\acronym} framework.
The cache configuration options are in Table~\ref{tab:config}.
We implement LRU, random \cite{lipp2016armageddon}, PLRU \cite{so1988cache}, and RRIP \cite{jaleel2010high} replacement policies in the cache simulator.
The cache simulator can be further extended to multi-level caches.
For simplicity, we currently use physically-indexed physically-tagged (PIPT) caches and let the attack and victim \lhl{programs}  directly use physical addresses for their accesses. 

\noindent{\bf Real hardware}:
we leverage CacheQuery~\cite{vila2020cachequery}, an open-source tool to directly measure the cache access timing on Intel processors. CacheQuery automatically figures out the address mappings for L1, L2 and L3 caches and provides access to a specific cache level, without disabling prefetching, turbo boost, frequency scaling, etc., which is close to real world operating conditions. 
Currently, CacheQuery~\cite{vila2020cachequery}  supports timing measurements for access sequences to one cache set. 
\\
Even though {\acronym} can interact with an arbitrary number of sets if the cache interface allows, our experiments focus on exploring attacks within small number of cache sets under different configurations, given that cache-timing attacks usually exploit the contention in each cache set independently.
%




\subsection{Attack and Victim \lhl{Program}  Configurations}

Cache-timing attacks also depend on how the memory space is shared between the attack program and the victim \lhl{program}, and which cache operations are available.
For example, if there is no shared memory between the victim \lhl{program} and the attack program, the flush+reload attack is not possible, but a prime+probe attack is still possible.
We refer to these choices as the attack and victim \lhl{program} configurations. 
In our setting, for the victim \lhl{program}'s access, the agent will not get the latency of that step (N.A. as the observation), \lhl{since we assume that the} victim \lhl{program}'s  access latency is not directly visible to the attack program.
To allow exploring attacks with and without shared memory space between the attack program and the victim \lhl{program}, the address range of the victim \lhl{program}  and  the address range of the attack program are configurable in {\acronym}, as listed in Table~\ref{tab:config}.
These address ranges determine whether the attack \lhl{program} can touch the same addresses accessed by the victim \lhl{program}.
In addition, cache-line flush instruction (\texttt{clflush} in x86) is not always available, {\em e.g.}, in JavaScript.  We make \texttt{flush\_enable} a configuration option in {\acronym}.

In many cache timing channels, rather than encoding the information with different addresses, whether the victim \lhl{program} has made access or not also leaks information, {\em e.g.}, prime+probe. 
To explore this scenario, we use the \texttt{victim\_no\_access\_enable} option, and use  $addr_{secret\_e}$ to represent the victim \lhl{program} makes no access when triggered.
When this is enabled, the victim \lhl{program}  can access one of the addresses in the victim \lhl{program}'s address space or make no access ($addr_{secret\_e}$) with the same probability.
To explore attacks under a detection scheme, we have the configuration option \texttt{detection\_enable}, which terminates an episode when the sequence is recognized as an attack by the detector. 
\subsection{RL Engine and Configuration}

\noindent{\bf RL Action Space.}
The RL agent can have the attack program take one of the three actions (i.e., $a_X$, $a_v$, $a_{gY}$), as discussed in Section~\ref{sec:rl_engine}.
If \texttt{flush\_enable} is set, we extend the action $a_X$ to also include cache line flush denoted by $a_{fX}$. 
Similarly, if \texttt{victim\_no\_access\_enable}, we use $a_{gE}$ to denote that the agent guesses that the victim program makes no access after triggered. When the attack program makes a guess, it can be either $a_{gY}$ or $a_{gE}$.
Overall, the RL agent can take one action for each step: 1) access/flush: $a_X$ or $a_{fX}$; 2)  trigger the victim \lhl{program}: $a_v$; 3) guess: $a_{gY}$  or $a_{gE}$, depending on the attack/victim \lhl{program}  configuration. 
We use one-hot encoding to represent the actions.

\begin{table*}[]
\vspace{+4pt}
\caption{Attack sequence found using \acronym~on real hardware.}\label{tab:cachequery}
\centering
\begin{tabular}{|@{}c@{}|@{}C{0.34in}@{}|@{}C{0.34in}@{}|@{}C{0.42in}@{}|@{}C{0.34in}@{}|@{}C{0.45in}@{}|@{}C{3.2in}@{}|@{}C{0.47in}@{}|@{}C{0.47in}@{}|}
 \hline
\bf CPU &\bf Cache level & \bf \#Ways &  \bf Rep. Pol. & \bf Victim addr. & \bf Attack addr.& \bf Example attack \hl{sequence} found by \acronym & \bf Accuracy & \bf Attack Category\\
\hline
\hline

\multicolumn{1}{|c|}{\multirow{3}{0.7in}{Core i7-6700 (SkyLake)}}   &L1 & 8& PLRU&0/E&\lhl{0-15} & \lhl{$10\rightarrow14 \rightarrow 13 \rightarrow12 \rightarrow 11 \rightarrow v \rightarrow 1 \rightarrow 3\rightarrow 6 \rightarrow 0 \rightarrow g$} & \lhl{1.0} & LRU\\
\cline{2-9}
&L2 & 4 & N.O.D.$^\ddagger$ &0/E&0-8 & \lhl{$8\rightarrow7\rightarrow4\rightarrow2\rightarrow v\rightarrow0\rightarrow g$}&\lhl{0.999}& LRU*\\
\cline{2-9}
&L3 & 4$^\dagger$ & N.O.D. &0/E&0-8 &\lhl{$1\rightarrow7\rightarrow 5\rightarrow6\rightarrow v\rightarrow0\rightarrow g$} & 1.0& LRU*\\ 
\cline{2-9}
\hline
\multicolumn{1}{|c|}{\multirow{2}{0.7in}{Core i7-7700K (KabyLake)}}
& L3 & 4$^\dagger$ & N.O.D.&0/E&0-8 & \lhl{$2\rightarrow4\rightarrow7\rightarrow8 \rightarrow4 \rightarrow4 \rightarrow v \rightarrow0 \rightarrow g $} &1.0& LRU*\\
\cline{2-9}
&L3 &8$^\dagger$& N.O.D.& 0/E& 0-15& \lhl{$15\rightarrow 11 \rightarrow 6\rightarrow 12\rightarrow 14 \rightarrow 8\rightarrow 9\rightarrow7\rightarrow 12 \rightarrow 2\rightarrow v\rightarrow 0\rightarrow1\rightarrow g$} & 0.993& LRU*\\
\hline
\multicolumn{1}{|c|}{\multirow{2}{0.7in}{\lhl{Core i7-9700 (CoffeeLake)}}}
& \lhl{L1} & 8 & PLRU &0/E& \lhl{0-15} & \lhl{$8\rightarrow 12 \rightarrow 3 \rightarrow 6 \rightarrow 15 \rightarrow 7\rightarrow 13 \rightarrow 4 \rightarrow 2 \rightarrow v \rightarrow 9 \rightarrow 0 \rightarrow g$}&\lhl{0.998}& LRU*\\
\cline{2-9}
&\lhl{L2} & 4 & N.O.D.& 0/E& 0-8&  \lhl{$3\rightarrow3\rightarrow3\rightarrow3\rightarrow8 \rightarrow2 \rightarrow6\rightarrow4 \rightarrow v \rightarrow v \rightarrow 0 \rightarrow g$}  & 1.0& LRU*\\
\hline

\end{tabular}
\begin{tablenotes}
   \item $\dagger$ indicates way partition using Intel CAT.  $\ddagger$ Not Officially Documented. * Attacks based on replacement states, but with different sequences due to the replacement policy.  
   \fhl{$\&$ We use non-distributed synchronous PPO \cite{schulman2017proximal} instead of asynchronous PPO for real hardware experiments due to less resource usage.}
  \end{tablenotes}

\vspace{-6pt}
\end{table*}

\noindent{\bf RL State Space.} For its own memory access $a_X$, the attack \lhl{program}  can directly observe the cache access latency in terms of a hit or a miss. 
To provide more information to let the RL agent learn efficiently,
we encode a state incorporating the history of actions and observations, and compose the state space $S$ as a Cartesian product of subspaces, as follows:
$     S = \Pi_{i=1}^W (S^i_{lat} \times S^i_{act} \times S^i_{step} \times S^i_{trig} )\label{eqn:state_space}$,
where $S^i_{lat}$, $S^i_{act}$,  $S^i_{step}$, $S^i_{trig}$ are the subspaces representing the access latency, an action taken, the current step, and whether the victim \lhl{program}  has already been triggered at step $i$. 
$W$ is the window size that can be set using \texttt{window\_size}. Empirically we set it to be 4-8 times \texttt{num\_blocks}. 
The latency subspace $S^i_{lat}$ is defined as $S^i_{lat}=\{s_{hit}$, $s_{miss}$, $s_{N.A.}\}$, representing hit/miss and N.A. states.
The action subspace $S^i_{act}$ is defined as $S^i_{act}= \{s_a |a\in \{ a_X,  a_{fX}, a_v, a_{gY}, a_{gE} \}\}$, representing the state in which action $a$ is taken at step $i$. 
The step subspace $S^i_{step}$ is defined as $S^i_{step} =\{s_{step1}, s_{step2}, ...\}$. 
The victim \lhl{program} triggering subspace $S^i_{trig}$ is defined as $S^i_{trig} =\{s_{t}, s_{nt}\}$, representing whether the victim \lhl{program} has been triggered or not. 
The RL agent uses this information in order to make a guess after the secret-dependent memory access by the victim \lhl{program} is triggered. 
\\
For real hardware, having the agent interact with hardware for each action is slow and also makes the training more susceptible to system noise. To address this challenge, for training on real hardware, we execute all instructions in an episode together as a batch. 
The latency is masked until the agent make a guess, when all instructions within an episode are executed and latency of memory accesses are revealed.

\noindent{\bf RL Algorithm and  DNN Model} 
In this paper, 
We use proximal policy optimization (PPO)~\cite{schulman2017proximal} since it typically performs similarly or better than other methods while being much simpler to tune \cite{openai}.  \chl{Our preliminary study also indicates that PPO converges faster than Ape-X DQN\cite{horgan2018distributed} in our context.} 
We use a Transformer \cite{vaswani2017attention} model (input feature dimensions 128, 1 layer encoder, 8-head, \lhl{feed-forward network} dimension 2048) as the backbone. Like BERT \cite{devlin-etal-2019-bert} model, we only use the Transformer Encoder to learn step-wise representation. We then use an average-pooling over steps to generate sequence embedding. 

\noindent{\bf Reward Configuration.}
\hl{When the agent makes a guess $a_{gY}$/$a_{gE}$, depending on whether the guess is correct or not, our environment assigns the reward \texttt{correct\_guess\_reward} or \texttt{wrong\_guess\_reward}.}
\hl{For all the actions taken by the RL agent (\textit{e.g.}, $a_X$, $a_{fX}$, $a_v$), the environment assigns a negative \texttt{step\_reward} to encourage the agent to find short attack sequences.} These rewards are listed in Table~\ref{tab:config}. 
PPO is not very sensitive to the reward value combinations and we use the following rewards for the experiments:
\fhl{\texttt{correct\_guess\_reward} = 1}, \fhl{\texttt{wrong\_guess\_reward} = -1}, and
\fhl{\texttt{step\_reward} = -0.01}. \lhl{For real hardware experiments, we use \fhl{\texttt{step\_reward} = -0.005} to explore longer sequences.} Once the sum of the reward within an episode is converged to a positive value, we use deterministic replay to extract the attack sequences.
\lhl{Tuning the reward values can  increase convergence speed for some specific cases but in general the reward values work for all the cases.}
 To discourage the RL agent from taking too many steps without making a guess, we have \texttt{length\_violation\_reward}, which is a large negative value when the length of an episode is longer than \texttt{window\_size}.
When we implement a detection scheme in the environment, there is also the \texttt{detection\_reward} which is the reward (negative value) when the sequence is caught by the detector as a potential attack. 




\subsection{Attack Analysis and Demonstration}

Once the attack sequence is generated using the RL engine, it can be further analyzed and demonstrated. First, for attack analysis, given a particular sequence, we want to know the type of the attack and whether it is a new type of attack. Currently, we rely on human inspections to classify  attacks and identify new attacks. Automated classification and identification of the attack sequences is an orthogonal problem and we left it as future work.
With the attack sequences generated by \acronym{}, we can demonstrate them in real hardware. We embed the attack sequence into an assembly template, which uses pointer chasing to perform measurement of timing of one access~\cite{xiong2021leaking}. With the assembly code corresponding to the attack sequence, we can then measure the bit rate and the error rate of the attack on a real processor with realistic noise.

\section{Evaluation and Case Studies}
\label{sec:case_studies}


In \acronym,
we use 
{RLMeta} \cite{rlmeta} as the RL framework to train the RL agent. 
The PPO implementation in RLMeta is the asynchronous PPO similar to Sample Factory \cite{petrenko2020sf}, with actors generating data and the learner learning asynchronously.
Our DNN model is implemented in  PyTorch~\cite{NEURIPS2019_9015}. The RL environment follows the OpenAI Gym~\cite{1606.01540} interface.
We implement the cache simulator based on \cite{cache_simulator}.
\lhl{Except for real hardware experiments whose hardwares are specified in the table, the training process is performed on clusters with Intel(R) Xeon(R) CPU E5-2698 v4 @ 2.20GHz CPUs and NVIDIA Tesla V100-SXM2-16GB GPUs.}

\subsection{Attacks on Real Hardware}

\begin{table*}[t]
\vspace{+4pt}
\centering
\caption{The RL environment configurations tested, and the example attack sequences generated by {\acronym}. }
\label{tab:action_space}
{\RaggedRight
\begin{tabular}{|@{ }c@{ }|c@{ }|@{ }c@{ }|@{ }c@{ }|c@{ }|@{ }c@{ }|@{ }c@{ }     |c@{ }|C{6.15cm}@{ }|@{}c@{ }|} 

\hline
    \multirow{2}{*}{\bf No.} &  \multicolumn{3}{c|}{\bf Cache config.} & \multicolumn{3}{c|}{\bf Attack\&victim config.} & \multicolumn{1}{c|}{\bf Expected attacks} & \multicolumn{2}{c|}{\bf Example Attack found by \acronym}\\
\cline{2-10}
  &Type$^\dagger$ & Ways & Sets & Victim   & Attack & Flush & Possible &  \multicolumn{1}{c|}{Attack sequence (p indicates prefetch)} & Attack \\ 
  &     & used       &       &  addr   & addr  &  inst &    attacks$^\ddagger$ &       & Category\\\hline

1 & DM & 1 & 4 & 0-3 &  4-7 & no  & PP &  \fhl{$ 7\rightarrow 4 \rightarrow 5 \rightarrow v \rightarrow 7\rightarrow 5 \rightarrow 4 \rightarrow g $} &  PP\\ \hline
  2 & DM+PFnextline & 1& 4 & 0-3   & 4-7 & no  & PP &  $6(p7)\rightarrow 7(p0) \rightarrow 5(p6) \rightarrow v \rightarrow 6(p7)\rightarrow 7(p0)\rightarrow 5(p6) \rightarrow g$ &  PP\\ \hline
 3& DM & 1 & 4 & 0-3 &  0-3 & yes & FR & \fhl{$ f0\rightarrow f3\rightarrow f2\rightarrow v\rightarrow2\rightarrow 3\rightarrow0\rightarrow g$}&FR\\ \hline

4& DM & 1 & 4 & 0-3 &  0-7 & no   & ER, PP & \fhl{$6\rightarrow 5\rightarrow 7\rightarrow v\rightarrow7 \rightarrow6\rightarrow1 \rightarrow g$} & ER and PP\\ \hline

 5& FA & 4 & 1 & 0/E &  4-7 & no  & PP, LRU & \fhl{$4\rightarrow5\rightarrow7\rightarrow v \rightarrow6\rightarrow4 \rightarrow g$} & LRU\\ \hline
 6& FA & 4 & 1 & 0/E &  0-3 & yes & FR, LRU & \fhl{$f0\rightarrow v\rightarrow0\rightarrow g$} &FR\\ \hline

  7& FA & 4 & 1 & 0/E &  0-7 & no  & ER, PP, LRU &  \fhl{$5\rightarrow 3 \rightarrow2\rightarrow1 \rightarrow v \rightarrow 0 \rightarrow g$} &LRU \\ \hline
 
  8& FA & 4 & 1 & 0-3 &  0-3 & yes & FR, LRU &  $f3\rightarrow f2\rightarrow f0\rightarrow v\rightarrow2\rightarrow3 \rightarrow 0 \rightarrow  g$ &FR \\  \hline
 9& FA & 4 & 1 & 0-3 &  0-7 & yes & FR, LRU &  $4\rightarrow f0\rightarrow 5 \rightarrow v \rightarrow 0 \rightarrow 2 \rightarrow 1 \rightarrow g$ & FR\\\hline
 10& DM & 1 & 8 & 0-7 &  0-7 & yes & FR & $f4\rightarrow f6\rightarrow f5\rightarrow f0\rightarrow f2\rightarrow f7\rightarrow v\rightarrow7\rightarrow6\rightarrow 5\rightarrow 2\rightarrow 4\rightarrow 0\rightarrow f1\rightarrow v\rightarrow 1\rightarrow  g$ &FR\\ \hline
 11& FA & 8 & 1 & 0/E &  0-7 & yes & FR, LRU & $f0\rightarrow v\rightarrow 0\rightarrow g$ &FR\\ \hline
 12& FA & 8 & 1 & 0/E &  0-15& no  & ER, PP, LRU &  \hl{$1\rightarrow13\rightarrow14\rightarrow15 \rightarrow5\rightarrow9\rightarrow11\rightarrow6\rightarrow v \rightarrow 0 \rightarrow g$} &ER\\ \hline
\hl{13} & \hl{FA+PFnextline} & \hl{8} & \hl{1} & \hl{0/E} &  \hl{0-15}& \hl{no}  & \hl{ER, PP, LRU} &  \hl{$1~(p2)\rightarrow 5~(p6)\rightarrow 10~(p11)\rightarrow8~(p9)\rightarrow 12~(p13)\rightarrow 13~(p14)\rightarrow 15~(p0)\rightarrow 4~(p5)\rightarrow v \rightarrow 0~(p1) \rightarrow g$} & \hl{ER} \\ \hline

\hl{14} & \hl{FA+PFstream} & \hl{8} & \hl{1} & \hl{0/E} & \hl{0-15}& \hl{no}  & \hl{ER, PP, LRU} &  \hl{$11\rightarrow 15\rightarrow7\rightarrow 4\rightarrow 6 \rightarrow 8(p10) \rightarrow1 \rightarrow 13\rightarrow v \rightarrow 0\rightarrow g$} & \hl{ER} \\ \hline

\chl{15}& \chl{SA} & \chl{2} & \chl{4} & \chl{0-3} &  \chl{4-11} & \chl{no}   & \chl{PP} & \chl{$4\rightarrow 7\rightarrow9 \rightarrow 8 \rightarrow 11 \rightarrow 5\rightarrow v\rightarrow 9 \rightarrow 7 \rightarrow4 \rightarrow g$} &  \chl{PP}\\ \hline
\chl{16}& \chl{2-level SA$^*$} & \chl{2$^*$} & \chl{4$^*$} & \chl{0-3} &  \chl{4-11} & \chl{no}   & \chl{PP} & \chl{$10\rightarrow11\rightarrow8 \rightarrow 7\rightarrow v\rightarrow 11\rightarrow 4\rightarrow8 \rightarrow6\rightarrow g$} &  \chl{PP}\\ \hline
\chl{17}& \chl{2-level SA$^*$} & \chl{2$^*$} & \chl{8$^*$} & \chl{0-7} &  \chl{8-23} & \chl{no}   & \chl{PP} & \chl{$12\rightarrow13\rightarrow 15\rightarrow17 \rightarrow 10\rightarrow 14\rightarrow21 \rightarrow 23\rightarrow 16 \rightarrow9 \rightarrow 18 \rightarrow v \rightarrow 15\rightarrow 13 \rightarrow10 \rightarrow 17 \rightarrow20\rightarrow12\rightarrow8\rightarrow16\rightarrow22 \rightarrow14\rightarrow g$} &  \chl{PP}\\ \hline


\end{tabular}
}
\begin{tablenotes}
   \item $\dagger$ FA: fully-associative,  DM:direct-mapped, \chl{SA: set-associative,}  PFnextline: nextline prefetcher, PFstream: stream prefetcher. 
   $\ddagger$ FR: flush+reload, ER: evict+reload, PP: prime+probe. \chl{0/E means the victim \lhl{program}  either accesses 0 or makes no access (i.e., {\bf E}mpty) when triggered. In the attack sequence, the number in the attack sequence is from the attack \lhl{program} address range, $v$ represents triggering victim \lhl{program} access, and $g$ represents making a guess. ($pn$) means prefetching $n$ which is done by the prefetcher automatically. $fn$ means flushing address $n$. $^*$2-core with 4-set-DM private L1 caches and a shared inclusive L2 cache where the victim \lhl{program}  and the attack \lhl{program}  each run on one core. The table shows the configuration of L2 cache.}
  \end{tablenotes}
\vspace{-12pt}
\end{table*}

\acronym~can explore attack sequences on real hardware without explicitly knowing all the architectural details, including associativity, replacement policies, frequency scaling, hardware prefetching, and other undocumented features.
In our experimental setup using CacheQuery~\cite{vila2020cachequery}, the attack program and the victim \lhl{program}  run in a single process on the same core. 
We experimented with multiple cache levels from three different processors,
all with the same attack and victim \lhl{program}  configurations 
where the attack program needs to guess whether the victim \lhl{program}   accesses the cache set or not. 
The configurations and the attack sequences found by \acronym{} are shown in Table~\ref{tab:cachequery}. 
The table also shows the accuracy for each attack sequence based on repeating the sequence 1,000 times on the same processor using CacheQuery. Due to noise in real processors, the accuracy is \lhl{slightly} less than 100\%. 

The results show that \acronym{} is able to find attack sequences on  real processors without explicitly specifying the number of ways or reverse engineering the replacement policies, prefetchers, etc. Such information is usually needed by human experts to adapt known attacks to a new platform. For example, to demonstrate prime+probe, one needs to understand the replacement policy to prime and probe a cache set efficiently~\cite{wang2019papp}. However, the replacement policies of recent processors are rarely documented publicly by the vendor and are difficult to precisely reverse engineer \cite{vila2020cachequery}. 
 Even though it is possible to reverse engineer an unspecified replacement policy from a real-world processor, it takes a significant amount of time (as long as 100 hours) and then one still needs to develop attack sequences manually. \acronym~can find effective attack sequences within {\em several~hours} in our experiments.


\subsection{Attacks on Diverse Cache/Attack Configurations} 

The flexibility of the cache simulator allows us to study diverse cache and attack configurations more easily.
To evaluate how effective the RL agent can be across a broad range of environments, we tested {\acronym} under many different cache and attack/victim \lhl{program}  configurations shown in Table~\ref{tab:action_space}. 
These environments use the (true) LRU replacement \lhl{policy} by default.
Note that the attack/victim \lhl{program}  configuration limits the feasible attacks in the environment. 
For example, if the environment does not allow the cache flush instruction, the flush+reload attack is not possible. 
If there is no shared address, flush+reload or evict+reload is not feasible. The expected attack category for each config are listed in Table~\ref{tab:action_space}. 




The RL agent can successfully find working attack sequences for all configurations we tested. Table~\ref{tab:action_space} shows one example attack sequence for each configuration that was automatically found by the RL agent. 
The RL-generated attack sequences vary for different environment configurations and cover a range of known attack categories, including prime+probe, flush+reload, and evict+reload.
\fhl{We use the Transformer model in {RLMeta} framework for all these configurations.}

In most cases, the RL agent generates attack sequences that are of the attack type expected for the configuration. 
Interestingly, the attack sequence found by the agent can be more efficient than the \chl{known attacks in the literature}.
For example, for configuration 1 in Table~\ref{tab:action_space}, a textbook prime+probe attack will result in the following attack\footnote{For brevity, we only use the subscript $i$ of an action $a_i$ to represent the the action. For example, to represent accessing address 3, we directly use $3$ to represent $a_3$.}:
$
    4\rightarrow 5 \rightarrow 6\rightarrow  7 \rightarrow v  \rightarrow 4\rightarrow 5 \rightarrow 6\rightarrow 7\rightarrow g
$.
Meanwhile, the attack sequence is given by \acronym~is: 
$7\rightarrow4\rightarrow5\rightarrow v\rightarrow7\rightarrow5 \rightarrow 4\rightarrow g$.
The RL agent  removes the unnecessary memory accesses in the textbook prime+probe attack, i.e., by attacking three cache sets, the attack \lhl{program} can already learn which of the four possible addresses the victim \lhl{program}  accessed. \chl{Most known attacks are written in for-loops for convenience, thus, it may contain more accesses than a solution RL found.}
 For configurations~5 and 7 in Table~\ref{tab:action_space}, instead of a prime+probe attack, the RL agent found a shorter attack sequence leveraging the LRU state. 
However, the attack sequences found by \acronym~are not always the shortest in length (e.g., configuration 10 in Table~\ref{tab:action_space}) and may contain unnecessary accesses. \chl{This is because the training converges to  local optima (whose guess accuracy is high but the length is not the shortest) instead of a global optimum, which might be hard to find in complex configurations}; nonetheless, they do capture the key mechanism that enables the attack for each configuration.
In some cases, the sequence found by the agent is an interesting combination of different attacks, e.g., configuration 4 in Table~\ref{tab:action_space} results in an attack that is a combination of evict+reload and prime+probe, which could make attack detection and defense more difficult.

\hl{The cache configurations 2, 13, and 14 include 
either a next-line prefetcher} \cite{smith1982cache} \hl{or a stream prefetcher} \cite{jouppi1990improving}. 
\hl{The RL agent could find attack sequences even with prefetching.} 

While not shown in the table for brevity, we also studied caches with a fixed random address-to-set mapping 
where an address is mapped to a set using a fixed random permutation. 
{\acronym} could find successful attacks after training on the cache with a given random permutation.



\subsection{Case Study 1: Attacking Replacement Policy}

\begin{table}[t]
\centering
\caption{RL training statistics and generated attacks for deterministic cache replacement policies.}
\label{tab:replacement_policies}
\begin{tabular}{|@{}C{0.4in}@{}|@{}C{0.5in}@{}|@{}C{0.45in}@{}|l|} 
 \hline
\bf Repl. alg. & \bf Epochs to converge$^{\$}$ &  \bf Episode length$^{\$}$ & \bf Attack sequence found by \acronym \\

 \hline
LRU &26.0& 7.0& \fhl{$3\rightarrow 1\rightarrow4 \rightarrow 2 \rightarrow v \rightarrow 0 \rightarrow g$} \\
\hline
PLRU &15.67&7.0& \fhl{$ 3 \rightarrow 4 \rightarrow 1 \rightarrow 2\rightarrow v \rightarrow 0 \rightarrow g$} \\
\hline 
RRIP &70.67& 12.7& \fhl{$2\rightarrow3\rightarrow3\rightarrow4 \rightarrow 1\rightarrow 4\rightarrow 2\rightarrow v\rightarrow 0 \rightarrow g$}\\
\hline 
\end{tabular}
\begin{tablenotes}
\item ${\$}$ Averaged over three training runs. One epoch is 3000 training steps.
\end{tablenotes}
\end{table}

\begin{table}[t]

\centering
\caption{\hl{The RL-generated attacks on the random replacement policy.}}
\label{tab:random_rep}

\begin{tabular}{|@{}C{1.1in}@{}|@{}C{1.1in}@{}|@{}C{1.1in}@{}|}
 \hline
\bf Step reward & \bf End accuracy  & \bf Episode length \\
\hline
-0.02& 0.98& 16.25\\
\hline 
-0.01&0.98& 18.85\\
\hline 
-0.005&0.94& 19.02\\
\hline 

\end{tabular}
\vspace{-12pt}
\end{table}

\acronym's cache simulator allows us to implement different replacement policies and address mappings in the same setting and the RL agent can adapt to them. We focus on replacement policy in this case study because recent attacks based on replacement states~\cite{xiong2020leaking,briongos2020reload+} show long and complex sequences, and we want to demonstrate the effectiveness of \acronym{}.
We use a 4-way cache (set) with four replacement policies: three deterministic (LRU, PLRU, and RRIP) and one non-deterministic (random). 
For a 4-way cache set, both LRU and RRIP keep 2-bit state information (ranging from 0-3)  for each cache block, called {\em age} in LRU and {\em re-reference prediction value (RRPV)} in RRIP, which will be incremented correspondingly. The cache block with the largest state bits will be evicted upon a cache miss. In LRU, the most recently used cache block will be assigned age=0. In RRIP, a newly installed cache block will be assigned RRPV=2, and only upon a cache hit will it be promoted to RRPV=0. Pseudo-LRU implemented using a tree structure is a way of approximating LRU with less state information, whose behavior will be slightly different from LRU.
The attack \lhl{program}'s address space is configured to be from 0 to 4 (large enough to fill the 4-way cache set).
The victim \lhl{program}  is configured to either access address 0 or make no access depending on a one-bit secret. The configuration is similar to that of configuration 6 in Table~\ref{tab:action_space}.

As shown in Table~\ref{tab:replacement_policies}, the RL agent can successfully generate valid attack sequences for all three deterministic policies, with the RL training time ranging from about 20 min to 3 hours.
 RRIP needs longer training time and a longer attack sequence compared to PLRU and LRU.
In the RRIP attack example, the 
$2\rightarrow3\rightarrow3\rightarrow4\rightarrow1\rightarrow4\rightarrow2 $
sequence is needed to ensure that 
RRPV is 0 for address 2, address 3 and address 4,  and address 0 will be evicted before the victim \lhl{program}  access. 
 For the deterministic replacement policies,  RL finds attacks that always make a correct guess, {\em i.e.}, there is no noise. 
Unlike a deterministic replacement policy where the next state will be fully determined given the action and current state, the next state is difficult to predict in the (pseudo)-random replacement policy. 
Thus, an attack sequence that results in a correct guess may result in a wrong guess in another evaluation. 
The RL agent can also produce different actions depending on the current observation. 
In that sense, unlike a deterministic replacement policy, there is no single attack sequence that always works in the random replacement policy, whose eviction rate depends on the number and the sequence of memory accesses.
Instead, we evaluate the attack accuracy of the RL agent over \lhl{100} evaluation runs. As shown in Table~\ref{tab:random_rep}, \hl{the step reward determines the tradeoff between the attack length and the accuracy.}\hl{ The evict+reload strategy is similar to the prior attack on random replacement policy}~\cite{lipp2016armageddon}.

\subsection{Case Study 2: Bypassing Defense and Detection Techniques}
\label{sec:bypass}



To protect against cache timing-channel attacks, a variety of detection and mitigation techniques have been proposed.
\acronym's cache simulator can be used to test research prototypes of defense and detection mechanisms proposed in previous works that lack actual real processor implementations. This is especially useful when evaluating new protection or detection schemes and finding vulnerabilities.
We implemented four cache timing-channel protection schemes in the cache simulator: 1) Partition-locked (PL) cache~\cite{wang2007new}, 2) Autocorrelation-based detector that is similar to CC-hunter~\cite{chen2014cc}, 3) machine learning-based detector that is similar to Cyclone~\cite{harris2019cyclone}, and 4) \hl{microarchitecture statistics}-based detection~\cite{zhang2016cloudradar,chiappetta2016real,alam2017performance,kulah2019spydetector},
{\acronym} successfully finds attack sequences that can bypass these protection schemes.

\noindent{\bf Partition-Locked (PL) Cache.}
PL cache~\cite{wang2007new} provides special instructions to lock specific cache lines in a cache to prevent them from being evicted. 
The victim program can lock its own cache lines so that they cannot be evicted by the attack program. Further, the victim \lhl{program}'s access to the locked cache lines will not evict any of the attack \lhl{program}'s cache lines. 
\hl{In}~\cite{he2017secure}\hl{, the formal analysis on a simplified cache model concludes that PL cache is secure when the attack \lhl{program} and the victim \lhl{program}  do not share address space.}

\begin{table}[t]
\vspace{+4pt}
\centering
\caption{Comparison of PLRU w/ and w/o PLCache.}
\label{tab:plcache}

\begin{tabular}{|@{}C{1.1in}@{}|@{}C{1.1in}@{}|@{}C{1.1in}@{}|} 
 \hline
 \bf Cache & \bf Epochs to converge$^{\$}$  & \bf Final episode length$^{\$}$   \\
 \hline
PL Cache &37.67& 8.1\\
\hline
Baseline &7.67&7.0 \\
\hline
\end{tabular}
\begin{tablenotes}
\item ${\$}$ Averaged over three training runs. One epoch is 3000 training steps.
\end{tablenotes}
\vspace{-12pt}
\end{table}

We implemented the PL cache with the lock/unlock interface in our cache model.
To use the PL cache as a defense mechanism, we assume the victim program's cache line is pre-installed and locked in the cache.  
We use a 4-way cache, and the address range of the attack \lhl{program}  is 1-5 and the victim \lhl{program}  either accesses 0 or has no access depending on the secret value. The setting is considered to be secure in~\cite{he2017secure}. 
Table~\ref{tab:plcache} shows the training time (in \# of epochs) and the attack sequence length for a cache with PLRU, with and without the PL cache.  The training runs take about an hour. 
{\acronym} successfully found an attack that works even with the PL cache, represented by attack sequence $1\rightarrow 3\rightarrow 5 \rightarrow 2 \rightarrow v\rightarrow 4 \rightarrow 4 \rightarrow g$.
In this attack, the victim program's cache line (address 0) always stays in the cache, and the victim program's behavior (whether the victim program makes access or not) does not evict any of the attack \lhl{program}'s cache lines. 
However, the victim program's access affects the LRU state. When the attack \lhl{program} made subsequent accesses, it can tell whether the victim program accessed address 0 or not by observing if a new block (e.g., address 4) can be brought into the cache.
This attack is reported in recent literature~\cite{xiong2020leaking}.


\noindent {\bf Autocorrelation-based Detection.} Autocorrelation of cache events has also been proposed to detect the existence of cache-timing channels \cite{chen2014cc,yan2016replayconfusion} based on the observation of the common covert-channel sequences. 
In CC-Hunter~\cite{chen2014cc}, \hl{two types of conflict miss events (i.e., the victim program evicting the attack program's cache line, $V\rightarrow A$, encoded with ``0'', and the attack \lhl{program} evicting the victim program's cache line, $A\rightarrow V$, encoded with ``1'')} are considered in the event train $\{X_i\}$ where $0\leq i\leq n$, and n is the length. 
In a contention-based cache side-channel attack like prime+probe, these two events are interleaved periodically. We can check the autocorrelation $C_p$ at lag $p$ using the following equation \cite{yan2016replayconfusion}:
$C_{p}=\frac{n\sum_{i=0}^{n-p}\left[\left(X_{i}-\bar{X}\right)\left(X_{i+p}-\bar{X}\right)\right]}{(n-p)\sum_{i=0}^{n}\left(X_{i}-\bar{X}\right)^{2}}$.
If there exists $p$ where $1\leq p\leq P$ 
($P$ is a predefined parameter) 
such that $C_p> C_{threshold}$ (e.g. 0.75), then it is an attack.

\hl{For example, in a 4-set direct-mapped cache where the victim and attack \lhl{program}'s address space is 0-3 and 4-7, a ``textbook'' prime+probe attack would perform the following steps. First, the address space 4-7 is  primed by the attack \lhl{program}, after that one victim program access is triggered, and then addresses 4-7 are probed. 
The event train and the corresponding  autocorrelogram is shown in Figure}~\ref{fig:autocorrelation}.\hl{ The maximum autocorrelation for $p\geq 1$ is $0.916$, which is over the threshold.}

\begin{figure}[t]
 \centering
 \includegraphics[width=0.95\columnwidth]{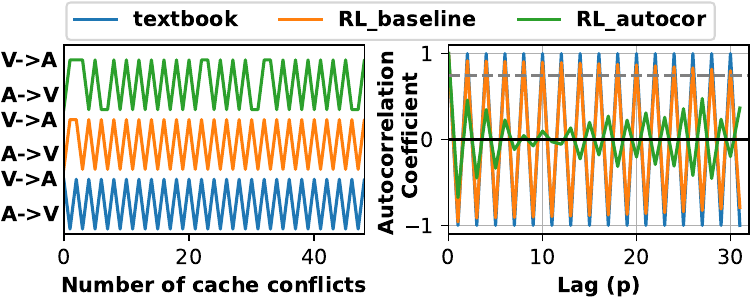}
 \caption{\hl{(a) Event train (A$\rightarrow$V: attack \lhl{program}'s conflict misses with the victim \lhl{program}  and V$\rightarrow$A: victim \lhl{program}'s conflict misses with the attack \lhl{program}.) (b) Autocorrelograms of the event trains. The dashed line shows the threshold for detecting an attack.}}
 \label{fig:autocorrelation}
 \end{figure}
 
\begin{table}[t]
\vspace{-6pt}
\centering
\caption{\hl{Bit rate, accuracy, and autocorrelation of attacks}.}
\vspace{-6pt}
\label{tbl:cchunter}
\renewcommand{\tabcolsep}{1.0mm} 
\begin{tabular}{|l|c@{}|c@{}|c@{}|}

\hline
\bf Attack&\bf Bit rate (guess/step)$^*$  &\bf Guess  accuracy$^*$ &\bf Avg max autocor$^*$ \\
\hline
Textbook & 0.1625   & 1.0 & 0.973 \\
\hline
RL\_baseline$^\$$ & \hl{0.229} &  \hl{0.989}&\hl{0.933} \\
\hline
RL\_autocor$^\$$ & \hl{0.216} & \hl{0.997} &\hl{0.519}   \\
\hline
\end{tabular}
\begin{tablenotes}
\item $\$$ Averaged over  three training runs.
\item $*$ Averaged over 1000 episodes. The last column shows the average of the maximum autocorrelation of each episode
\end{tablenotes}
\vspace{-12pt}
\end{table}

\lhl{AutoCAT can train a baseline attack agent where multiple guesses happen in one fixed-step (e.g.,  160-step) episode and each guess corresponds to one secret. The more correct guesses it makes, the higher reward it will get. There is a negative reward when there is no guess action in the episode to encourage guesses.}
\hl{However, the AutoCAT generated baseline sequence (\textit{RL\_baseline}) can be also detected by this autocorrelation-based detector, as shown in the conflict miss event train and the autocorrelogram} (Figure~\ref{fig:autocorrelation}). \hl{The maximum autocorrelation for $p\geq1$ is $0.916$. Note that the event train and the corresponding autocorrelograms will be different for each sample. }
However, AutoCAT can learn to bypass the autocorrelation-based detector if the reward of the RL agent is augmented to avoid high autocorrelation.
\hl{We use $L_2$-penalty of ${C_p}$ to penalize high autocorrelation, which is defined as}
$R_{L_2} = a  \sum^P_{p=1} \frac{C_p^2}{P}$
\hl{where $a$ is a negative number and $P<< n$ is the length of ${C_p}$ used for autocorrelation-based detection.}
The sampled cache conflict miss event train and the autocorrelogram of the resulting agent (\textit{RL\_autocor}) are shown in Figure~\ref{fig:autocorrelation}. The maximum autocorrelation beyond lag 0 is $0.477$ in this example. 
The results indicate that the agent (\textit{RL\_autocor}) will be able to evade autocorrelagoram detection with high probability.

\hl{Table~}\ref{tbl:cchunter} \hl{compares the attack sequence from the textbook, \textit{RL\_baseline}, and \textit{RL\_autocor} in terms of average bit rate (number of guesses per step), average accuracy, and average maximum autocorrelation of each episode. 
All attacks achieve over $0.98$ accuracy and the RL agents have a higher bit rate than the textbook attack. This is because the RL agents optimize the bit rate to gain higher rewards. 
We observe when a miss is already observed during a probe step, RL agents can guess the secret while the textbook attack still completes the remaining accesses. }
We also observe that the bit rate of \textit{RL\_autocor} is lower than \textit{RL\_baseline} because \textit{RL\_autocor} makes additional accesses to reduce autocorrelation.

\noindent {\bf ML-based Detection.}
\hl{Machine learning classifiers can also be used for detecting cache-timing attacks. For example, in Cyclone}~\cite{harris2019cyclone}, the frequency of cyclic access sequences  by different security domains  (e.g., $ a\leadsto b\leadsto a$) for each cache line within each time interval is used as the input of an SVM classifier to detect cache timing channels efficiently.
We use a 4-set direct-mapped cache as an example and implement domain tracking and cyclic access sequence counting for each cache line following \cite{harris2019cyclone}\lhl{. We train an SVM-based detector using SPEC2017 benchmarks for benign memory access traces and the textbook prime+probe attack for malicious memory traces. The 5-fold validation accuracy of the SVM detector in correctly predicting the benign/malicious traces is 98.8\%.} 

\hl{We then train the AutoCAT's RL agent (}{\em RL\_SVM}) \hl{with this detector}. \hl{If the SVM detector correctly reports the existence of an attack, the RL agent gets a negative reward. We also train an} {\em RL\_baseline} \hl{agent without detection penalty. }
\hl{We show the result in Table}~\ref{tbl:cyclone}. \hl{The textbook and }{\em RL\_baseline} attacks \hl{can be easily detected by the SVM detector, with the detection rate of 0.997 and 0.715, respectively. However, when the RL agent is trained with the SVM detection penalty, it can find attack sequences that can bypass the SVM detector, with the detection rate of 0.00333, at the cost of a reduced bit rate. This indicates ML-based detector trained on static traces can be bypassed by an RL-based attack and a novel training method for ML-based detector is needed. }

\begin{table}[t]
\vspace{+4pt}
\centering
\caption{\hl{Comparison of bit rate, guess accuracy, and detection rate by the SVM}.}
\label{tbl:cyclone}
\renewcommand{\tabcolsep}{1.0mm} 
\begin{tabular}{|l|c|c|c|}
\hline
\bf attacker &\bf bit rate (guess/step)& \bf  guess accuracy &\bf detection rate\\
\hline
textbook & \hl{0.1625} &  \hl{1.0} & \hl{0.997} \\
\hline
RL\_baseline ${^\$}$ & \hl{0.228} & \hl{0.998} & \hl{0.715}${^*}$ \\
\hline
RL\_SVM $^{\$}$ & \hl{0.168} & \hl{0.998} & \hl{0.00333} \\
\hline
\end{tabular}
\begin{tablenotes}
\item ${\$}$ averaged over  three training runs. ranging from 0.338 to 0.909.
\end{tablenotes}
\end{table}

\noindent{\bf \hl{$\mu$arch Statistics}-based Detection.}
Most of the cache-timing attacks cause the victim \lhl{program}'s process to incur more cache misses during the attack. 
Thus, \hl{detection schemes based on microarchitecture event counts/statistics} have been proposed to leverage hardware performance counters (HPCs) to monitor the cache hit-rate of the victim \lhl{program}  process and detect an attack at run-time~\cite{zhang2016cloudradar,chiappetta2016real,alam2017performance,kulah2019spydetector}.
\hl{More recently,~}\cite{mirbagher2020perspectron}\hl{ observes that fine-grained statistics can achieve better detection accuracy.}
When an abnormally large number of cache misses are observed, the detector signals a potential attack.

\begin{figure}[t]
\centering
\includegraphics[width=0.99\columnwidth]{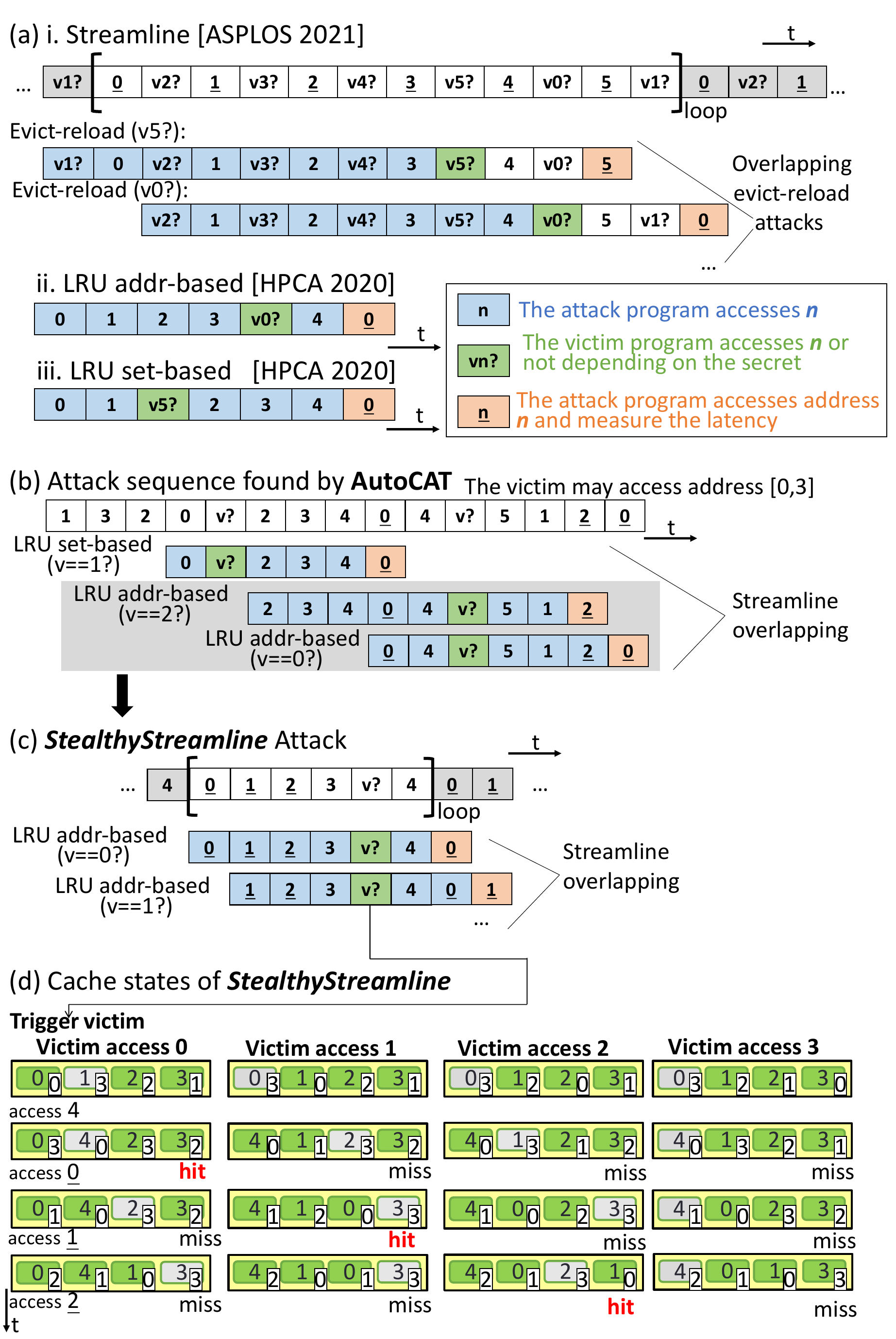}
\caption{StealthyStreamline attack. (a) Known attack sequences in the literature. (b) The new attack sequence found by \acronym, which can be seen as the combination of the two known attacks. (c) StealthyStreamline attack derived from the attack found by \acronym.
(d) The cache state changes of the StealthyStreamline attack. The numbers in the right bottom corner indicate the LRU age of the cache line.}
\label{fig:Stealthystreamline_diagram}
\vspace{-12pt}
\end{figure}

To evaluate the statistics-based detection scheme in \acronym, we consider an attack is detected when the victim \lhl{program}'s access triggers a cache miss. 
We terminate the episode and assign a negative reward if the victim \lhl{program}'s access results in its cache miss during training.
This configuration encourages the RL agent to avoid victim \lhl{program}'s misses, and thus, avoid the miss-based detection.

Figure~\ref{fig:Stealthystreamline_diagram}(b) shows the attack sequence generated by {\acronym} for a 4-way cache with the miss-based detection scheme. 
This is a new attack sequence, which is a novel combination of the two recent attacks in literature (shown in  Figure~\ref{fig:Stealthystreamline_diagram}(a)). 
The attack sequence can be divided into sub-sequences, which are the LRU set-based or LRU address-based attacks \cite{xiong2020leaking}. 
The two sub-sequences overlap with each other in a way similar to the Streamline attack~\cite{saileshwar2021streamline}. 
Based on the gray part of the sequence generated by {\acronym} in Figure~\ref{fig:Stealthystreamline_diagram}(b), we construct a new attack, named \textit{StealthyStreamline}, in Figure~\ref{fig:Stealthystreamline_diagram}(c).
Compared to the Streamline attack, the new StealthyStreamline attack does not cause cache misses of the victim \lhl{program} and thus is stealthier. 
Compared to the LRU-based attacks, StealthyStreamline has a higher bit rate by overlapping the steps for multiple bits, effectively transferring multiple bits at a time. 
Figure~\ref{fig:Stealthystreamline_diagram}(d) illustrates the cache state of StealthyStreamline. The attack program observes different timing when the victim program holds different secret values.

\subsection{Demonstration of Attacks on Real Machines}
\label{sec:real_attack}

The attack sequence discovered by \acronym{} captures one essential aspect of real-world cache-timing attacks. However, real-world attacks also have to deal with common practical issues like measurement noise, calibrations, interferences, etc. These practical issues are common in many attacks and people handle them in a similar way regardless of the actual attack sequences. To demonstrate the \lhl{effectiveness} of these attack sequences in a real-world scenario, we put the attack sequence from {\acronym} into an open-source attack assembly template \cite{xiong2021leaking}, which handles calibration, measurement,  cache line access, and forms a covert channel corresponding to the attack sequence. We can then execute the assembly on a real processor under a practical operating environment.


With this method, we \lhl{tested} the \textit{StealthyStreamline} attack sequence for a covert channel \hl{in the L1 data cache on four different Intel processors (Table}~\ref{tbl:real_attack}\hl{). Two processors have a 32KB (8-way) L1 data cache, and the other two latest processors employ a 48KB (12-way) L1 data cache. The machines run different versions of Linux.} 

\hl{We generalize the 2-bit \textit{StealthyStreamline} attack sequence in a 4-way cache (in Figure}~\ref{fig:Stealthystreamline_diagram}) \hl{to 8-way and 12-way scenarios by adding extra accesses to the cache lines that map to the same cache set. We implemented both 2-bit (4 possible $addr_{secret}$ values) and 3-bit (8 possible $addr_{secret}$ values) \textit{StealthyStreamline} covert channels.}

\begin{table}[t]
\vspace{+4pt}
\centering
\caption{\hl{Covert channels on real machines.}}
\label{tbl:real_attack}
 \begin{threeparttable}
\begin{tabular}{|C{0.4in}|@{ } c@{ }|@{ } c@{ }|@{ } c@{ }||@{ }c@{ }|@{ }c@{ }|@{ }c@{ }|} 
 \hline
 \multirow{2}{*}{\bf CPU} & \multirow{2}{*}{\bf $\mu$-arch.} & \multirow{2}{*}{\bf L1D config} 
 & \multirow{2}{*}{\bf OS} 
 & \multicolumn{3}{@{ }c@{ }|}{\bf{Bit Rate$^a$~ (Mbps)}} \\
  \cline{5-7}
 & & & & {\bf LRU}& {\bf SS.$^b$~} & {\bf Impr.}\\
  \hline
\hl{Xeon E5- 2687Wv2} &  \hl{IvyBridge}& \hl{32KB(8way)}& \hl{Ubuntu18} & \hl{6.2} &\hl{ 7.7} &\hl{$24\%$ }\\
\hline
\hl{Core i7-6700} & \hl{Skylake} & \hl{32KB(8way)}&\hl{ Ubuntu18} & \hl{3.6} & \hl{4.5} &\hl{$22\%$}\\
\hline 
\hl{Core i5-11600K} & \hl{RocketLake} & \hl{48KB(12way)}& \hl{CentOS8}& \hl{3.4  }& \hl{5.7 }&\hl{$67\%$} \\
\hline
\hl{Xeon W-1350P} & \hl{RocketLake} & \hl{48KB(12way)} & \hl{Ubuntu20 }& \hl{2.1} &\hl{3.7} &\hl{$71\%$}\\
\hline
\end{tabular}
\begin{tablenotes}
   \item $^a$ The bit rate when the average error rate $<$ 5\%. $^b$SS. for StealthyStreamline.  
  \end{tablenotes}
 \end{threeparttable}
\vspace{-12pt}
\end{table}

\begin{figure}[t]
\centering
\includegraphics[width=0.99\columnwidth]{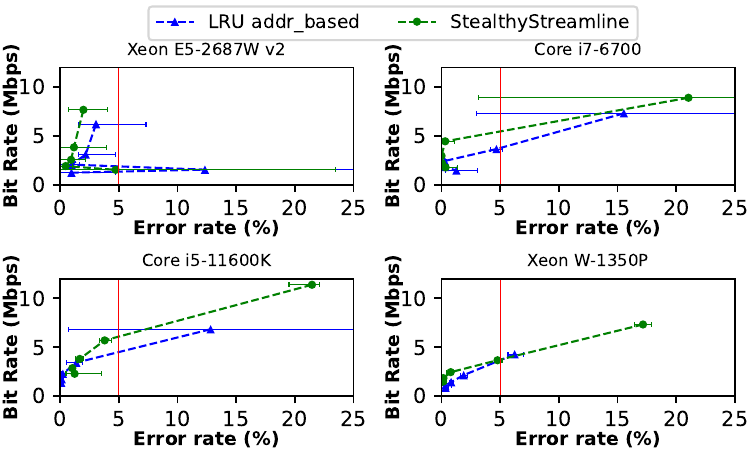}
\caption{Bit rate of the StealthyStreamline attack and the LRU address-based attack \hl{on four different processors}. The horizontal error bars show the range of errors across different transmission runs.}
\label{fig:Stealthystreamline_error}
\end{figure}

\noindent\textbf{Bit Rate and Error Rate:}
We test the bit rate and the corresponding error rate of the covert channel within a process.  
\hl{We do not change any system configuration with the purpose of facilitating the attack, {\em i.e.}, hardware prefetchers remain enabled.}
We measure the bit rate by measuring the time of sending a 2048-bit random string 100 times, using \texttt{time} in Linux. 
We evaluate the error rate using the Hamming Distance between the message being sent and the message received. 
 We observe that the 3-bit \textit{StealthyStreamline} has a high error rate due to the tree structure in PLRU, while the 2-bit \textit{StealthyStreamline} has a low error rate.

Figure~\ref{fig:Stealthystreamline_error} shows the bit rates and the corresponding error rates for the \hl{2-bit} \textit{StealthyStreamline} covert channel and the baseline LRU address-based covert channel. 
\hl{In most of the machines,} we observe the LRU address-based covert channel has a larger variation in the error rate across different experiment runs, as shown by the error bars in Figure~\ref{fig:Stealthystreamline_error}. 
For the error rate less than $5\%$, StealthyStreamline has a higher bit rate than the LRU address-based covert channel. 
\hl{In an 8-way cache, StealthyStreamline has up to a $24\%$ higher bit rate.
In the latest 12-way cache,} the StealthyStreamline has up to a $71\%$ higher bit rate. \hl{StealthyStreamline improves the bit rate more for caches with higher associativity because a smaller fraction of memory accesses (4 out of 10 for the 8-way cache vs. 4 out of 14 for the 12-way cache) need to be measured and measuring the latency of access takes more cycles than normal memory accesses.}


\noindent\textbf{Spectre Attack using StealthyStreamline:} \hl{We \lhl{test} Spectre V1 attack}\cite{kocher2019spectre}\hl{ with the StealthyStreamline as the covert channel. Compared with the LRU address-based covert channel, the 2-bit StealthyStreamline enables us to encode 4x more symbols with the same cache. Compared with flush+reload or evict+time, StealthyStreamline makes the attack stealthier.}

\section{Discussion and Future Work}
\label{sec:discussions}
\subsection{Comparison with Search Algorithms}

RL takes fewer steps to find a successful attack than a brute-force search and has the potential to handle much larger search spaces. 
Consider the prime+probe attack on an $N$-way cache set as an example. 
On average, we can find one prime+probe sequence every $M$ sequences, where
    $M=\frac{2\times (N+1)^{2N+1}}{(N!)^2} $.
Since $N!\sim \sqrt{2\pi N}(\frac{N}{e})^{2N}$, we have $M\sim e^{2N}$, which increases exponentially with the number of ways. 
For $N=8$, $M \approx 2.05\times 10^7$, it takes about 369 million steps to find an attack, considering each attack sequence takes $2N+2$ steps. Last-level caches usually have more than 8 ways and it will be infeasible for an exhaustive search. With RL, the agent converges within $\sim$1 million steps.

\hl{Compared to RL which learns policy/value on the fly to progressively improve the search quality to find the distinguishing sequence, traditional search techniques may not have sufficient learning capability. For example, a random search does not have learning at all; A* search utilizes a predefined heuristics function, which was not updated during the search process.}

\subsection{Factors Affecting Training Speed}

Several factors, including ML model architecture and the initialization of the RL environment, affect the training speed and convergence. While we mainly report the results from Transformer models, we also tested other models such as multi-layer perceptron (MLP). MLP is capable of finding attacks for small configurations, but typically takes more time to converge. Also, in our experiments, the cache is initialized by accessing a sequence of addresses randomly sampled from the attack and victim programs' address ranges. Different initialization schemes may change the time to converge during training. We leave finding an optimal initialization scheme for training speed as future work.

\subsection{Future Extensions}

\hl{One challenge in AutoCAT lies in analyzing many attack sequences produced by an RL agent. In this study, we manually analyzed and categorized the sequences found by AutoCAT. Ideally, we want to automate the attack analysis.}
\chl{For example, we plan to develop a classification model to decide the type of attacks found by \acronym.} 

\hl{Given the repetitive structure of a cache, we focus on training AutoCAT with small cache configurations and generalize the attacks to real machines manually. The scalability and generalizability when applying to multiple levels of large caches still need to be further explored.}
Our initial investigation shows promising results that suggest AutoCAT can be further improved 
with RL generalization methods~\cite{raileanu_automatic_2021,kirk2021genrlsurvey}. 
\hl{In future work, we plan to investigate extending the AutoCAT approach to more complex cache configurations and other types of microarchitectural timing channels beyond caches.}

For training on real hardware, the CacheQuery interface only supports accesses to one cache set, which limits studying attack sequences that involve multiple cache sets. We plan to make CacheQuery support multiple cache~sets.
\section{Related Work}
\label{sec:related_work}

\noindent{\bf Automated timing channel analysis and discovery.}
Prior studies proposed systematic analysis methods for cache timing channels \hl{based on simplified cache models}~\cite{he2017secure,deng2020benchmark,xiao2019speechminer}. 
Our work demonstrates RL as a new way to enable more automated security analysis, 
which can be easily extended to new systems or defense mechanisms with less human effort.
There exist other automated approaches for vulnerability analysis such as exhaustive approach~\cite{fadiheh2022exhaustive}, taint analysis~\cite{gras2020absynthe}, fuzzing~\cite{weber2021osiris, moghimi2020medusa,ghaniyoun2021introspectre}, relational testing \cite{oleksenko2022revizor}, and formal methods \cite{trippel2018checkmate}. 
CheckMate~\cite{trippel2018checkmate}  can only discover attacks that is specified by the given exploit sequence.
Information flow tracking technique such as SecVerilog~\cite{zhang2015hardware} can identify information leakage in the designs but cannot generate exploits automatically.
For example, the attack sequences that exploit dirty bits \cite{cui2022abusing} are only reported seven years after the vulnerability is noted by SecVerilog.
Compared to formal methods, the RL-based method cannot provide mathematical security guarantees, but can more easily be applied to a system without building a formal model or writing proofs, both of which require significant human efforts \cite{buiras2021validation}.
Compared to fuzzing, the RL-based method is more efficient and expressive for complex attack sequences. 
For example, Osiris~\cite{weber2021osiris} explores  attack with three instructions, and IntroSpectre~\cite{ghaniyoun2021introspectre} relies on predefined attack gadgets. 


\noindent{\bf Cache-timing attack detection and defense.}
To prevent cache timing-channel attacks, many detection and defense mechanisms were proposed. Detection mechanisms such as cc-Hunter \cite{chen2014cc} and ReplayConfusion \cite{yan2016replayconfusion} focus on detecting an attack. 
We show that the RL agent has the potential to automatically generate attacks to evade detection schemes.  
On the other hand, defenses
\cite{yan2017secure,tan2020phantomcache,dessouky2020hybcache,saileshwar2021mirage,qureshi2019new,ojha2021timecache, kiriansky2018dawg}
focus on mitigating or removing the interference that leads to known timing channels. 
This paper shows that RL also has the potential to automatically evaluate the security of mitigations, and shows that {\acronym} can  break PL cache \cite{wang2007new} successfully. 


\noindent{\bf ML for security.}
Machine learning was used in the computer security domain for anomaly detection \cite{lane2000machine}, website fingerprinting \cite{la2021wireless,shusterman2019robust}, and other analysis tasks. 
However, traditional supervised learning cannot find new attacks without known attack sequences or labels.
To address this challenge, we propose to use RL, which can be trained with delayed rewards and whose action trajectories are expressive enough to represent real-world attack sequences.

Reinforcement learning has been used for software security \cite{nguyen2019deep},
IoT security \cite{uprety2020reinforcement},
autonomous driving security \cite{rasheed2020deep},
power side-channel attacks \cite{ramezanpour2020scarl},
circuit test generation \cite{pan2020test},
power side-channel countermeasures \cite{cryptoeprint:2021:526}, and
hardware Trojan detection \cite{10.1145/3394885.3431595}. 
To the best of our knowledge, our work is the first of its kind in using RL to actively and automatically generate attack sequences in the microarchitecture security~domain.

\section{Conclusion}
\label{sec:conclusion}

In this paper, we propose to use reinforcement learning to automatically find existing and undiscovered timing-channel attack sequences. 
As a concrete example, we build the {\acronym} framework, which can explore cache-timing attacks in various cache configurations and attack/victim \lhl{program} settings, and under different defense and detection mechanisms.
Our experimental results show that the RL agent can find practical attack sequences for various blackbox cache designs. 
The RL agent also discovered the StealthyStreamline attack, which is a novel attack with a higher bit rate on real machines than attacks reported in previous literature. 
{\acronym} shows RL is a promising method to explore microarchitecture timing attacks in practical systems. 


\section*{Acknowledgment}

This project is partially supported by NSF grant ECCS-1932501 and Commonwealth Cybersecurity Initiative. We thank the anonymous reviewers 
for their constructive feedback. 
We thank John~G.~Harris at Virginia Tech for technical support and Yifang~Liu at Cornell University for providing a computer for experiments. 
We thank  Chris Cummins, Hugh Leather, Andrew~C.~Myers, Benjamin~C.~Lee, Jiaxun~Cui, and Yanqi~Zhang for helpful discussions.




\bibliographystyle{IEEEtranS}
\bibliography{refs}

\begin{thebibliography}{10}
\providecommand{\url}[1]{#1}
\csname url@samestyle\endcsname
\providecommand{\newblock}{\relax}
\providecommand{\bibinfo}[2]{#2}
\providecommand{\BIBentrySTDinterwordspacing}{\spaceskip=0pt\relax}
\providecommand{\BIBentryALTinterwordstretchfactor}{4}
\providecommand{\BIBentryALTinterwordspacing}{\spaceskip=\fontdimen2\font plus
\BIBentryALTinterwordstretchfactor\fontdimen3\font minus
  \fontdimen4\font\relax}
\providecommand{\BIBforeignlanguage}[2]{{%
\expandafter\ifx\csname l@#1\endcsname\relax
\typeout{** WARNING: IEEEtranS.bst: No hyphenation pattern has been}%
\typeout{** loaded for the language `#1'. Using the pattern for}%
\typeout{** the default language instead.}%
\else
\language=\csname l@#1\endcsname
\fi
#2}}
\providecommand{\BIBdecl}{\relax}
\BIBdecl

\bibitem{cache_simulator}
\BIBentryALTinterwordspacing
 [Online]. Available: \url{https://github.com/auxiliary/CacheSimulator}
\BIBentrySTDinterwordspacing

\bibitem{openai}
\BIBentryALTinterwordspacing
 [Online]. Available: \url{https://openai.com/blog/openai-baselines-ppo}
\BIBentrySTDinterwordspacing

\bibitem{abel2014reverse}
A.~Abel and J.~Reineke, ``Reverse engineering of cache replacement policies in
  intel microprocessors and their evaluation,'' in \emph{2014 IEEE
  International Symposium on Performance Analysis of Systems and Software
  (ISPASS)}.\hskip 1em plus 0.5em minus 0.4em\relax IEEE, 2014, pp. 141--142.

\bibitem{alam2017performance}
M.~Alam, S.~Bhattacharya, D.~Mukhopadhyay, and S.~Bhattacharya, ``Performance
  counters to rescue: A machine learning based safeguard against
  micro-architectural side-channel-attacks,'' \emph{Cryptology ePrint Archive},
  2017.

\bibitem{berner2019dota}
C.~Berner, G.~Brockman, B.~Chan, V.~Cheung, P.~Debiak, C.~Dennison, D.~Farhi,
  Q.~Fischer, S.~Hashme, C.~Hesse \emph{et~al.}, ``Dota 2 with large scale deep
  reinforcement learning,'' \emph{arXiv preprint arXiv:1912.06680}, 2019.

\bibitem{bernsteincache}
D.~J. Bernstein, ``Cache-timing attacks on {AES},'' 2005.

\bibitem{bonneau2006cache}
J.~Bonneau and I.~Mironov, ``Cache-collision timing attacks against aes,'' in
  \emph{International Workshop on Cryptographic Hardware and Embedded
  Systems}.\hskip 1em plus 0.5em minus 0.4em\relax Springer, 2006, pp.
  201--215.

\bibitem{briongos2020reload+}
S.~Briongos, P.~Malag{\'o}n, J.~M. Moya, and T.~Eisenbarth, ``{RELOAD+
  REFRESH}: Abusing cache replacement policies to perform stealthy cache
  attacks,'' in \emph{29th USENIX Security Symposium (USENIX Security 20)},
  2020, pp. 1967--1984.

\bibitem{1606.01540}
G.~Brockman, V.~Cheung, L.~Pettersson, J.~Schneider, J.~Schulman, J.~Tang, and
  W.~Zaremba, ``Open{AI} {Gym},'' 2016.

\bibitem{buiras2021validation}
P.~Buiras, H.~Nemati, A.~Lindner, and R.~Guanciale, ``Validation of
  side-channel models via observation refinement,'' in \emph{MICRO-54: 54th
  Annual IEEE/ACM International Symposium on Microarchitecture}, 2021, pp.
  578--591.

\bibitem{chen2014cc}
J.~Chen and G.~Venkataramani, ``{CC}-hunter: Uncovering covert timing channels
  on shared processor hardware,'' in \emph{2014 47th Annual IEEE/ACM
  International Symposium on Microarchitecture}.\hskip 1em plus 0.5em minus
  0.4em\relax IEEE, 2014, pp. 216--228.

\bibitem{chen2021leaking}
Y.~Chen, L.~Pei, and T.~E. Carlson, ``Leaking control flow information via the
  hardware prefetcher,'' \emph{arXiv preprint arXiv:2109.00474}, 2021.

\bibitem{chiappetta2016real}
M.~Chiappetta, E.~Savas, and C.~Yilmaz, ``Real time detection of cache-based
  side-channel attacks using hardware performance counters,'' \emph{Applied
  Soft Computing}, vol.~49, pp. 1162--1174, 2016.

\bibitem{cui2022abusing}
Y.~Cui and X.~Cheng, ``Abusing cache line dirty states to leak information in
  commercial processors,'' in \emph{2022 IEEE International Symposium on High
  Performance Computer Architecture (HPCA)}.\hskip 1em plus 0.5em minus
  0.4em\relax IEEE, 2022.

\bibitem{deng2020benchmark}
S.~Deng, W.~Xiong, and J.~Szefer, ``A benchmark suite for evaluating caches'
  vulnerability to timing attacks,'' in \emph{Proceedings of the Twenty-Fifth
  International Conference on Architectural Support for Programming Languages
  and Operating Systems}, 2020, pp. 683--697.

\bibitem{dessouky2020hybcache}
G.~Dessouky, T.~Frassetto, and A.-R. Sadeghi, ``Hyb{C}ache: Hybrid
  side-channel-resilient caches for trusted execution environments,'' in
  \emph{29th USENIX Security Symposium (USENIX Security 20)}, 2020, pp.
  451--468.

\bibitem{devlin-etal-2019-bert}
\BIBentryALTinterwordspacing
J.~Devlin, M.-W. Chang, K.~Lee, and K.~Toutanova, ``{BERT}: Pre-training of
  deep bidirectional transformers for language understanding,'' in
  \emph{Proceedings of the 2019 Conference of the North {A}merican Chapter of
  the Association for Computational Linguistics: Human Language Technologies,
  Volume 1 (Long and Short Papers)}.\hskip 1em plus 0.5em minus 0.4em\relax
  Minneapolis, Minnesota: Association for Computational Linguistics, Jun. 2019,
  pp. 4171--4186. [Online]. Available: \url{https://aclanthology.org/N19-1423}
\BIBentrySTDinterwordspacing

\bibitem{fadiheh2022exhaustive}
M.~R. Fadiheh, A.~Wezel, J.~Muller, J.~Bormann, S.~Ray, J.~M. Fung, S.~Mitra,
  D.~Stoffel, and W.~Kunz, ``An exhaustive approach to detecting transient
  execution side channels in {RTL} designs of processors,'' \emph{IEEE
  Transactions on Computers}, 2022.

\bibitem{ghaniyoun2021introspectre}
M.~Ghaniyoun, K.~Barber, Y.~Zhang, and R.~Teodorescu, ``{IntroSpectre}: a
  pre-silicon framework for discovery and analysis of transient execution
  vulnerabilities,'' in \emph{2021 ACM/IEEE 48th Annual International Symposium
  on Computer Architecture (ISCA)}.\hskip 1em plus 0.5em minus 0.4em\relax
  IEEE, 2021, pp. 874--887.

\bibitem{gras2020absynthe}
B.~Gras, C.~Giuffrida, M.~Kurth, H.~Bos, and K.~Razavi, ``{ABSynthe}: Automatic
  blackbox side-channel synthesis on commodity microarchitectures.'' in
  \emph{Network and Distributed Systems Security (NDSS) Symposium}, 2020.

\bibitem{gruss2016flush+}
D.~Gruss, C.~Maurice, K.~Wagner, and S.~Mangard, ``{Flush+ Flush}: a fast and
  stealthy cache attack,'' in \emph{International Conference on Detection of
  Intrusions and Malware, and Vulnerability Assessment}.\hskip 1em plus 0.5em
  minus 0.4em\relax Springer, 2016, pp. 279--299.

\bibitem{harris2019cyclone}
A.~Harris, S.~Wei, P.~Sahu, P.~Kumar, T.~Austin, and M.~Tiwari, ``Cyclone:
  Detecting contention-based cache information leaks through cyclic
  interference,'' in \emph{Proceedings of the 52nd Annual IEEE/ACM
  International Symposium on Microarchitecture}, 2019, pp. 57--72.

\bibitem{he2017secure}
Z.~He and R.~B. Lee, ``How secure is your cache against side-channel attacks?''
  in \emph{Proceedings of the 50th Annual IEEE/ACM International Symposium on
  Microarchitecture}, 2017, pp. 341--353.

\bibitem{horgan2018distributed}
D.~Horgan, J.~Quan, D.~Budden, G.~Barth-Maron, M.~Hessel, H.~van Hasselt, and
  D.~Silver, ``Distributed prioritized experience replay,'' in
  \emph{International Conference on Learning Representations}, 2018.

\bibitem{hsiao2021synthesizing}
Y.~Hsiao, D.~P. Mulligan, N.~Nikoleris, G.~Petri, and C.~Trippel,
  ``Synthesizing formal models of hardware from {RTL} for efficient
  verification of memory model implementations,'' in \emph{MICRO-54: 54th
  Annual IEEE/ACM International Symposium on Microarchitecture}, 2021, pp.
  679--694.

\bibitem{jaleel2010high}
A.~Jaleel, K.~B. Theobald, S.~C. Steely~Jr, and J.~Emer, ``{High performance
  cache replacement using re-reference interval prediction (RRIP)},'' \emph{ACM
  SIGARCH Computer Architecture News}, vol.~38, no.~3, pp. 60--71, 2010.

\bibitem{jouppi1990improving}
N.~P. Jouppi, ``Improving direct-mapped cache performance by the addition of a
  small fully-associative cache and prefetch buffers,'' \emph{ACM SIGARCH
  Computer Architecture News}, vol.~18, no. 2SI, pp. 364--373, 1990.

\bibitem{kiriansky2018dawg}
V.~Kiriansky, I.~Lebedev, S.~Amarasinghe, S.~Devadas, and J.~Emer, ``{DAWG}: A
  defense against cache timing attacks in speculative execution processors,''
  in \emph{2018 51st Annual IEEE/ACM International Symposium on
  Microarchitecture (MICRO)}.\hskip 1em plus 0.5em minus 0.4em\relax IEEE,
  2018, pp. 974--987.

\bibitem{kirk2021genrlsurvey}
R.~Kirk, A.~Zhang, E.~Grefenstette, and T.~Rockt{\"{a}}schel, ``A survey of
  generalisation in deep reinforcement learning,'' \emph{CoRR}, vol.
  abs/2111.09794, 2021.

\bibitem{kocher2019spectre}
P.~Kocher, J.~Horn, A.~Fogh, D.~Genkin, D.~Gruss, W.~Haas, M.~Hamburg, M.~Lipp,
  S.~Mangard, T.~Prescher \emph{et~al.}, ``Spectre attacks: Exploiting
  speculative execution,'' in \emph{2019 IEEE Symposium on Security and Privacy
  (SP)}.\hskip 1em plus 0.5em minus 0.4em\relax IEEE, 2019, pp. 1--19.

\bibitem{kulah2019spydetector}
Y.~Kulah, B.~Dincer, C.~Yilmaz, and E.~Savas, ``{SpyDetector}: An approach for
  detecting side-channel attacks at runtime,'' \emph{International Journal of
  Information Security}, vol.~18, no.~4, pp. 393--422, 2019.

\bibitem{la2021wireless}
A.~S. La~Cour, K.~K. Afridi, and G.~E. Suh, ``Wireless charging power
  side-channel attacks,'' in \emph{Proceedings of the 2021 ACM SIGSAC
  Conference on Computer and Communications Security}, 2021, pp. 651--665.

\bibitem{lane2000machine}
T.~D. Lane, \emph{Machine learning techniques for the computer security domain
  of anomaly detection}.\hskip 1em plus 0.5em minus 0.4em\relax Purdue
  University, 2000.

\bibitem{lazaric2012transfer}
A.~Lazaric, ``Transfer in reinforcement learning: a framework and a survey,''
  in \emph{Reinforcement Learning}.\hskip 1em plus 0.5em minus 0.4em\relax
  Springer, 2012, pp. 143--173.

\bibitem{lipp2016armageddon}
M.~Lipp, D.~Gruss, R.~Spreitzer, C.~Maurice, and S.~Mangard, ``{ARM}ageddon:
  Cache attacks on mobile devices,'' in \emph{25th USENIX Security Symposium
  (USENIX Security 16)}, 2016, pp. 549--564.

\bibitem{lipp2018meltdown}
M.~Lipp, M.~Schwarz, D.~Gruss, T.~Prescher, W.~Haas, A.~Fogh, J.~Horn,
  S.~Mangard, P.~Kocher, D.~Genkin \emph{et~al.}, ``Meltdown: Reading kernel
  memory from user space,'' in \emph{27th USENIX Security Symposium (USENIX
  Security 18)}, 2018, pp. 973--990.

\bibitem{liu2015last}
F.~Liu, Y.~Yarom, Q.~Ge, G.~Heiser, and R.~B. Lee, ``Last-level cache
  side-channel attacks are practical,'' in \emph{2015 IEEE symposium on
  security and privacy}.\hskip 1em plus 0.5em minus 0.4em\relax IEEE, 2015, pp.
  605--622.

\bibitem{luo2020stealthy}
M.~Luo, A.~C. Myers, and G.~E. Suh, ``Stealthy tracking of autonomous vehicles
  with cache side channels,'' in \emph{29th USENIX Security Symposium (USENIX
  Security 20)}.\hskip 1em plus 0.5em minus 0.4em\relax USENIX Association,
  Aug. 2020, pp. 859--876.

\bibitem{michalevsky2015powerspy}
Y.~Michalevsky, A.~Schulman, G.~A. Veerapandian, D.~Boneh, and G.~Nakibly,
  ``{PowerSpy}: Location tracking using mobile device power analysis,'' in
  \emph{24th USENIX Security Symposium (USENIX Security 15)}, 2015, pp.
  785--800.

\bibitem{mirbagher2020perspectron}
S.~Mirbagher-Ajorpaz, G.~Pokam, E.~Mohammadian-Koruyeh, E.~Garza,
  N.~Abu-Ghazaleh, and D.~A. Jim{\'e}nez, ``Perspectron: Detecting invariant
  footprints of microarchitectural attacks with perceptron,'' in \emph{2020
  53rd Annual IEEE/ACM International Symposium on Microarchitecture
  (MICRO)}.\hskip 1em plus 0.5em minus 0.4em\relax IEEE, 2020, pp. 1124--1137.

\bibitem{mnih_human-level_2015}
V.~Mnih, K.~Kavukcuoglu, D.~Silver, A.~A. Rusu, J.~Veness, M.~G. Bellemare,
  A.~Graves, M.~Riedmiller, A.~K. Fidjeland, G.~Ostrovski, S.~Petersen,
  C.~Beattie, A.~Sadik, I.~Antonoglou, H.~King, D.~Kumaran, D.~Wierstra,
  S.~Legg, and D.~Hassabis, ``Human-level control through deep reinforcement
  learning,'' \emph{Nature}, vol. 518, no. 7540, pp. 529--533, Feb. 2015,
  publisher: Nature Publishing Group, a division of Macmillan Publishers
  Limited. All Rights Reserved.

\bibitem{moghimi2020medusa}
D.~Moghimi, M.~Lipp, B.~Sunar, and M.~Schwarz, ``Medusa: Microarchitectural
  data leakage via automated attack synthesis,'' in \emph{29th USENIX Security
  Symposium (USENIX Security 20)}, 2020, pp. 1427--1444.

\bibitem{narvekar2020curriculum}
S.~Narvekar, B.~Peng, M.~Leonetti, J.~Sinapov, M.~E. Taylor, and P.~Stone,
  ``Curriculum learning for reinforcement learning domains: A framework and
  survey,'' \emph{arXiv preprint arXiv:2003.04960}, 2020.

\bibitem{nguyen2019deep}
T.~T. Nguyen and V.~J. Reddi, ``Deep reinforcement learning for cyber
  security,'' \emph{IEEE Transactions on Neural Networks and Learning Systems},
  2019.

\bibitem{ojha2021timecache}
D.~Ojha and S.~Dwarkadas, ``Time{Cache}: using time to eliminate cache side
  channels when sharing software,'' in \emph{2021 ACM/IEEE 48th Annual
  International Symposium on Computer Architecture (ISCA)}.\hskip 1em plus
  0.5em minus 0.4em\relax IEEE, 2021, pp. 375--387.

\bibitem{oleksenko2022revizor}
O.~Oleksenko, C.~Fetzer, B.~K{\"o}pf, and M.~Silberstein, ``Revizor: testing
  black-box cpus against speculation contracts,'' in \emph{Proceedings of the
  27th ACM International Conference on Architectural Support for Programming
  Languages and Operating Systems}, 2022, pp. 226--239.

\bibitem{osvik2006cache}
D.~A. Osvik, A.~Shamir, and E.~Tromer, ``Cache attacks and countermeasures: the
  case of {AES},'' in \emph{Cryptographers’ track at the {RSA}
  conference}.\hskip 1em plus 0.5em minus 0.4em\relax Springer, 2006, pp.
  1--20.

\bibitem{10.1145/3394885.3431595}
Z.~Pan and P.~Mishra, ``Automated test generation for hardware trojan detection
  using reinforcement learning.''\hskip 1em plus 0.5em minus 0.4em\relax New
  York, NY, USA: Association for Computing Machinery, 2021.

\bibitem{pan2020test}
Z.~Pan, J.~Sheldon, and P.~Mishra, ``Test generation using reinforcement
  learning for delay-based side-channel analysis,'' in \emph{2020 IEEE/ACM
  International Conference On Computer Aided Design (ICCAD)}.\hskip 1em plus
  0.5em minus 0.4em\relax IEEE, 2020, pp. 1--7.

\bibitem{NEURIPS2019_9015}
A.~Paszke, S.~Gross, F.~Massa, A.~Lerer, J.~Bradbury, G.~Chanan, T.~Killeen,
  Z.~Lin, N.~Gimelshein, L.~Antiga, A.~Desmaison, A.~Kopf, E.~Yang, Z.~DeVito,
  M.~Raison, A.~Tejani, S.~Chilamkurthy, B.~Steiner, L.~Fang, J.~Bai, and
  S.~Chintala, ``{PyTorch}: An imperative style, high-performance deep learning
  library,'' in \emph{Advances in Neural Information Processing Systems 32},
  H.~Wallach, H.~Larochelle, A.~Beygelzimer, F.~d\textquotesingle
  Alch\'{e}-Buc, E.~Fox, and R.~Garnett, Eds.\hskip 1em plus 0.5em minus
  0.4em\relax Curran Associates, Inc., 2019, pp. 8024--8035.

\bibitem{perez2011functional}
W.~Perez, E.~Sanchez, M.~S. Reorda, A.~Tonda, and J.~V. Medina, ``Functional
  test generation for the plru replacement mechanism of embedded cache
  memories,'' in \emph{2011 12th Latin American Test Workshop (LATW)}.\hskip
  1em plus 0.5em minus 0.4em\relax IEEE, 2011, pp. 1--6.

\bibitem{petrenko2020sf}
\BIBentryALTinterwordspacing
A.~Petrenko, Z.~Huang, T.~Kumar, G.~Sukhatme, and V.~Koltun, ``Sample factory:
  Egocentric 3{D} control from pixels at 100000 {FPS} with asynchronous
  reinforcement learning,'' in \emph{Proceedings of the 37th International
  Conference on Machine Learning}, ser. Proceedings of Machine Learning
  Research, H.~D. III and A.~Singh, Eds., vol. 119.\hskip 1em plus 0.5em minus
  0.4em\relax PMLR, 13--18 Jul 2020, pp. 7652--7662. [Online]. Available:
  \url{https://proceedings.mlr.press/v119/petrenko20a.html}
\BIBentrySTDinterwordspacing

\bibitem{qureshi2019new}
M.~K. Qureshi, ``New attacks and defense for encrypted-address cache,'' in
  \emph{2019 ACM/IEEE 46th Annual International Symposium on Computer
  Architecture (ISCA)}.\hskip 1em plus 0.5em minus 0.4em\relax IEEE, 2019, pp.
  360--371.

\bibitem{raileanu_automatic_2021}
R.~Raileanu, M.~Goldstein, D.~Yarats, I.~Kostrikov, and R.~Fergus, ``Automatic
  {Data} {Augmentation} for {Generalization} in {Reinforcement} {Learning},''
  in \emph{Advances in {Neural} {Information} {Processing} {Systems}},
  M.~Ranzato, A.~Beygelzimer, Y.~Dauphin, P.~S. Liang, and J.~W. Vaughan, Eds.,
  vol.~34.\hskip 1em plus 0.5em minus 0.4em\relax Curran Associates, Inc.,
  2021, pp. 5402--5415.

\bibitem{ramezanpour2020scarl}
K.~Ramezanpour, P.~Ampadu, and W.~Diehl, ``{SCARL}: side-channel analysis with
  reinforcement learning on the ascon authenticated cipher,'' \emph{arXiv
  preprint arXiv:2006.03995}, 2020.

\bibitem{rasheed2020deep}
I.~Rasheed, F.~Hu, and L.~Zhang, ``Deep reinforcement learning approach for
  autonomous vehicle systems for maintaining security and safety using
  {LSTM-GAN},'' \emph{Vehicular Communications}, vol.~26, p. 100266, 2020.

\bibitem{cryptoeprint:2021:526}
J.~Rijsdijk, L.~Wu, and G.~Perin, ``Reinforcement learning-based design of
  side-channel countermeasures,'' Cryptology ePrint Archive, Report 2021/526,
  2021.

\bibitem{saileshwar2021streamline}
G.~Saileshwar, C.~W. Fletcher, and M.~Qureshi, ``Streamline: a fast, flushless
  cache covert-channel attack by enabling asynchronous collusion,'' in
  \emph{Proceedings of the 26th ACM International Conference on Architectural
  Support for Programming Languages and Operating Systems}, 2021, pp.
  1077--1090.

\bibitem{saileshwar2021mirage}
G.~Saileshwar and M.~Qureshi, ``{MIRAGE}: Mitigating conflict-based cache
  attacks with a practical fully-associative design,'' in \emph{30th USENIX
  Security Symposium (USENIX Security 21)}, 2021, pp. 1379--1396.

\bibitem{schulman2017proximal}
J.~Schulman, F.~Wolski, P.~Dhariwal, A.~Radford, and O.~Klimov, ``Proximal
  policy optimization algorithms,'' \emph{arXiv preprint arXiv:1707.06347},
  2017.

\bibitem{shusterman2019robust}
A.~Shusterman, L.~Kang, Y.~Haskal, Y.~Meltser, P.~Mittal, Y.~Oren, and
  Y.~Yarom, ``Robust website fingerprinting through the cache occupancy
  channel,'' in \emph{28th USENIX Security Symposium (USENIX Security 19)},
  2019, pp. 639--656.

\bibitem{silver_mastering_2016}
D.~Silver, A.~Huang, C.~J. Maddison, A.~Guez, L.~Sifre, G.~van~den Driessche,
  J.~Schrittwieser, I.~Antonoglou, V.~Panneershelvam, M.~Lanctot, S.~Dieleman,
  D.~Grewe, J.~Nham, N.~Kalchbrenner, I.~Sutskever, T.~Lillicrap, M.~Leach,
  K.~Kavukcuoglu, T.~Graepel, and D.~Hassabis, ``Mastering the {Game} of {Go}
  with {Deep} {Neural} {Networks} and {Tree} {Search},'' \emph{Nature}, vol.
  529, no. 7587, pp. 484--489, Jan. 2016.

\bibitem{silver2017mastering}
D.~Silver, J.~Schrittwieser, K.~Simonyan, I.~Antonoglou, A.~Huang, A.~Guez,
  T.~Hubert, L.~Baker, M.~Lai, A.~Bolton \emph{et~al.}, ``Mastering the game of
  go without human knowledge,'' \emph{nature}, vol. 550, no. 7676, pp.
  354--359, 2017.

\bibitem{smith1982cache}
A.~J. Smith, ``Cache memories,'' \emph{ACM Computing Surveys (CSUR)}, vol.~14,
  no.~3, pp. 473--530, 1982.

\bibitem{so1988cache}
K.~So and R.~N. Rechtschaffen, ``Cache operations by {MRU} change,'' \emph{IEEE
  Transactions on Computers}, vol.~37, no.~6, pp. 700--709, 1988.

\bibitem{tan2020phantomcache}
Q.~Tan, Z.~Zeng, K.~Bu, and K.~Ren, ``{PhantomCache}: Obfuscating cache
  conflicts with localized randomization.'' in \emph{Networked and Distributed
  System Symposium (NDSS)}, 2020.

\bibitem{trippel2018checkmate}
C.~Trippel, D.~Lustig, and M.~Martonosi, ``Check{M}ate: Automated synthesis of
  hardware exploits and security litmus tests,'' in \emph{2018 51st Annual
  IEEE/ACM International Symposium on Microarchitecture (MICRO)}.\hskip 1em
  plus 0.5em minus 0.4em\relax IEEE, 2018, pp. 947--960.

\bibitem{uprety2020reinforcement}
A.~Uprety and D.~B. Rawat, ``Reinforcement learning for {IoT} security: A
  comprehensive survey,'' \emph{IEEE Internet of Things Journal}, vol.~8,
  no.~11, pp. 8693--8706, 2020.

\bibitem{vaswani2017attention}
A.~Vaswani, N.~Shazeer, N.~Parmar, J.~Uszkoreit, L.~Jones, A.~N. Gomez,
  L.~Kaiser, and I.~Polosukhin, ``Attention is all you need,'' \emph{Advances
  in neural information processing systems}, vol.~30, 2017.

\bibitem{vila2020cachequery}
P.~Vila, P.~Ganty, M.~Guarnieri, and B.~K{\"o}pf, ``{CacheQuery}: Learning
  replacement policies from hardware caches,'' in \emph{Proceedings of the 41st
  ACM SIGPLAN Conference on Programming Language Design and Implementation},
  2020, pp. 519--532.

\bibitem{wang2019papp}
D.~Wang, Z.~Qian, N.~Abu-Ghazaleh, and S.~V. Krishnamurthy, ``{PAPP}:
  Prefetcher-aware prime and probe side-channel attack,'' in \emph{Proceedings
  of the 56th Annual Design Automation Conference 2019}, 2019, pp. 1--6.

\bibitem{wang2007new}
Z.~Wang and R.~B. Lee, ``New cache designs for thwarting software cache-based
  side channel attacks,'' in \emph{Proceedings of the 34th annual international
  symposium on Computer architecture}, 2007, pp. 494--505.

\bibitem{weber2021osiris}
D.~Weber, A.~Ibrahim, H.~Nemati, M.~Schwarz, and C.~Rossow, ``Osiris: Automated
  discovery of microarchitectural side channels,'' in \emph{30th USENIX
  Security Symposium (USENIX Security 21)}, 2021, pp. 1415--1432.

\bibitem{wei2018know}
L.~Wei, B.~Luo, Y.~Li, Y.~Liu, and Q.~Xu, ``I know what you see: Power
  side-channel attack on convolutional neural network accelerators,'' in
  \emph{Proceedings of the 34th Annual Computer Security Applications
  Conference}, 2018, pp. 393--406.

\bibitem{xiao2019speechminer}
Y.~Xiao, Y.~Zhang, and R.~Teodorescu, ``{SPEECHMINER}: A framework for
  investigating and measuring speculative execution vulnerabilities,''
  \emph{arXiv preprint arXiv:1912.00329}, 2019.

\bibitem{xiong2021leaking}
W.~Xiong, S.~Katzenbeisser, and J.~Szefer, ``Leaking information through cache
  {LRU} states in commercial processors and secure caches,'' \emph{IEEE
  Transactions on Computers}, vol.~70, no.~4, pp. 511--523, 2021.

\bibitem{xiong2020leaking}
W.~Xiong and J.~Szefer, ``Leaking information through cache {LRU} states,'' in
  \emph{2020 IEEE International Symposium on High Performance Computer
  Architecture (HPCA)}.\hskip 1em plus 0.5em minus 0.4em\relax IEEE, 2020, pp.
  139--152.

\bibitem{yan2017secure}
M.~Yan, B.~Gopireddy, T.~Shull, and J.~Torrellas, ``Secure hierarchy-aware
  cache replacement policy ({SHARP}): Defending against cache-based side
  channel attacks,'' in \emph{2017 ACM/IEEE 44th Annual International Symposium
  on Computer Architecture (ISCA)}.\hskip 1em plus 0.5em minus 0.4em\relax
  IEEE, 2017, pp. 347--360.

\bibitem{yan2016replayconfusion}
M.~Yan, Y.~Shalabi, and J.~Torrellas, ``{ReplayConfusion}: detecting
  cache-based covert channel attacks using record and replay,'' in \emph{2016
  49th Annual IEEE/ACM International Symposium on Microarchitecture
  (MICRO)}.\hskip 1em plus 0.5em minus 0.4em\relax IEEE, 2016, pp. 1--14.

\bibitem{yan2019attack}
M.~Yan, R.~Sprabery, B.~Gopireddy, C.~Fletcher, R.~Campbell, and J.~Torrellas,
  ``Attack directories, not caches: Side channel attacks in a non-inclusive
  world,'' in \emph{2019 IEEE Symposium on Security and Privacy (SP)}.\hskip
  1em plus 0.5em minus 0.4em\relax IEEE, 2019, pp. 888--904.

\bibitem{rlmeta}
\BIBentryALTinterwordspacing
X.~Yang, B.~Cui, T.~Li, and Y.~Tian, ``{RLMeta: A Flexible Framework for
  Distributed Reinforcement Learning},'' 1 2022. [Online]. Available:
  \url{https://github.com/facebookresearch/rlmeta}
\BIBentrySTDinterwordspacing

\bibitem{yao2018coherence}
F.~Yao, M.~Doroslovacki, and G.~Venkataramani, ``Are coherence protocol states
  vulnerable to information leakage?'' in \emph{2018 IEEE International
  Symposium on High Performance Computer Architecture (HPCA)}.\hskip 1em plus
  0.5em minus 0.4em\relax IEEE, 2018, pp. 168--179.

\bibitem{yarom2014flush+}
Y.~Yarom and K.~Falkner, ``{FLUSH+ RELOAD}: A high resolution, low noise, l3
  cache side-channel attack,'' in \emph{23rd USENIX security symposium (USENIX
  security 14)}, 2014, pp. 719--732.

\bibitem{yuan2022automated}
Y.~Yuan, Q.~Pang, and S.~Wang, ``Automated side channel analysis of media
  software with manifold learning,'' in \emph{31st USENIX Security Symposium
  (USENIX Security 22)}.\hskip 1em plus 0.5em minus 0.4em\relax Boston, MA:
  USENIX Association, Aug. 2022.

\bibitem{zhang2015hardware}
D.~Zhang, Y.~Wang, G.~E. Suh, and A.~C. Myers, ``A hardware design language for
  timing-sensitive information-flow security,'' \emph{ASPLOS '15}, p.
  503–516, 2015.

\bibitem{zhang2016cloudradar}
T.~Zhang, Y.~Zhang, and R.~B. Lee, ``Cloud{R}adar: A real-time side-channel
  attack detection system in clouds,'' in \emph{International Symposium on
  Research in Attacks, Intrusions, and Defenses}.\hskip 1em plus 0.5em minus
  0.4em\relax Springer, 2016, pp. 118--140.

\end{thebibliography}
\appendix
\section{Artifact Appendix}

\subsection{Abstract}


Our artifact contains the AutoCAT framework (a cache simulator + scripts for launching the training and evaluation), as well as code for performing the StealthyStreamline attack on real processors.

\subsection{Artifact check-list (meta-information)}


\noindent{\bf AutoCAT}
{\small
\begin{itemize}
  \item {\bf Algorithm: } PPO training cache gueesing game 
  \item {\bf Run-time environment: } Linux 
  \item {\bf Hardware: } multi-core CPU with NVIDIA GPUs (with CUDA support)
  \item {\bf Disk space required: } About 10GB
  \item {\bf Time needed to prepare workflow: } About 1hr
  \item {\bf Time needed to complete experiments: } About 3hr (using checkpoints).
  \item {\bf Code licenses: } GPLv2
\end{itemize}
}

\noindent{\bf StealthyStreamline Attack}

{\small
\begin{itemize}
  \item {\bf Algorithm: } StealthyStreamline cache timing channel attack
  \item {\bf Run-time environment: } Linux
  \item {\bf Hardware: } Intel x86 CPU
  \item {\bf Metrics: } Bandwidth and error rate
  \item {\bf Disk space required: } About 1GB
  \item {\bf Time needed to prepare workflow: } About 10 mins
  \item {\bf Time needed to complete experiments: } About 10 mins
\end{itemize}
}


\subsection{Description}

Here is a brief description of the artifact. You can find more details in the artifact README file.

\subsubsection{How to access}
The code is available at:

\url{https://github.com/facebookresearch/AutoCAT}.




\subsubsection{Hardware dependencies}

The cache simulator training requires one physical machine with NVIDIA GPUs.
We use Intel(R) Xeon(R) CPU E5-2687W v2 @ 3.40GHz with NVIDIA Tesla K80 GPUs.

\subsubsection{Software dependencies}
\begin{itemize}
    \item Linux 18.04 or higher 
    \item Python 3.8 
    \item Pytorch 1.12
    \item CUDA 10.2 or higher
    \item gcc 9.3
    \item conda environment
\end{itemize}



\subsection{Installation}

In the Linux terminal, the following command downloads the artifact:
\begin{lstlisting}
$ git clone https://github.com/facebookresearch/AutoCAT
\end{lstlisting}

To experiment with the artifact, some dependencies need to be installed manually.
We use Conda to manage all the Python dependencies, we assume Conda is already installed.
Creating a conda environment:

\begin{lstlisting}
$ conda create --name py38 python=3.8
\end{lstlisting}

Then press enter when prompt.

Activate the conda environment
\begin{lstlisting}
$ conda activate py38
\end{lstlisting}

Under the py38 environment, install PyTorch as the following. (note that different machines may have different instructions on how to install PyTorch, please follow \url{https://pytorch.org/get-started/previous-versions/} for specific machines. The following command works on the aforementioned setup.)

\begin{lstlisting}
(py38) $ conda install pytorch==1.12.1 torchvision==0.13.1 torchaudio==0.12.1 cudatoolkit=10.2 -c pytorch
\end{lstlisting}

Install gcc9.3 (which supports C++17, required for building moolib).

\begin{lstlisting}
(py38) $ conda install gcc_linux-64=9.3.0
(py38) $ conda install gxx_linux-64=9.3.0
\end{lstlisting}
Install and build moolib.

\begin{lstlisting}
(py38) $ pip install scikit-learn seaborn pyyaml hydra-core terminaltables pep517
(py38) $ pip install git+https://github.com/facebookresearch/moolib.git@06e7a3e80c9f52729b4a6159f3fb4fc78986c98e
\end{lstlisting}

The environment is based on OpenAI Gym. To install it, use the following.

\begin{lstlisting}
$ pip install gym==0.26
\end{lstlisting}

The RL trainer is based on RLMeta.

Please follow setup process on rlmeta for installing RLMeta.

\begin{lstlisting}
$ git clone https://github.com/facebookresearch/rlmeta
$ cd rlmeta
$ git checkout 1057fbbf2637a002296afe5071e6ac0e7b630fe0
$ git submodule sync
$ git submodule update --init --recursive
$ pip install -e .
\end{lstlisting}

These should install all the dependencies needed.



\subsection{Evaluation and expected results}

\begin{enumerate} 

\item {\bf Table IV: attacks patterns found on CacheSimulator}

We expect different configurations to have different attack patterns as shown in Table IV.

\item {\bf Table V: RL training with different replacement policies}

We expect the RRIP policy to take more steps/epochs to converge than LRU/PLRU.

\item {\bf Table VI: random replacement policies}

We expect larger step rewards to result in lower guessing accuracy.

\item {\bf Table VII: comparison of PLRU with and without PLCache}

We expect training for PLCache to take more steps/epochs.

\item {\bf Table VIII: bit rate, autocorrection and accuracy for attacks bypassing CC-Hunter}

We expect textbook and RL\_baseline to have high max autocorrelation, while RL\_autocor to have low max autocorrelation. On the other hand, RL\_autocor's bit rate is lower.

\item {\bf Table IX: bit rate, guessing accuracy, and detection rate for attacks bypassing SVM-based detection}

We expect texbook and RL\_baseline to have a high SVM detection rate, while RL\_SVM to have a low SVM detection rate.

\item {\bf Figure 4: measuring the bit rate and the error rate for the SteathyStreamline attack}

We expect the bit rate of StealthyStreamline to be higher than that of the LRU addr\_based attack when the error rate is low (5\%).

\end{enumerate}


\subsection{Methodology}

Submission, reviewing and badging methodology:

\begin{itemize}
  \item \url{https://www.acm.org/publications/policies/artifact-review-badging}
  \item \url{http://cTuning.org/ae/submission-20201122.html}
  \item \url{http://cTuning.org/ae/reviewing-20201122.html}
\end{itemize}



\end{document}